\documentclass[fleqn,usenatbib]{aa}
\usepackage[utf8]{inputenc}
\usepackage{comment}

\usepackage{graphicx}
\usepackage{amsmath}
\usepackage{nccmath}
\usepackage{natbib} 
\usepackage{amssymb}
\usepackage{txfonts}
\usepackage[rightcaption]{sidecap}
\usepackage[hidelinks]{hyperref}

\usepackage{newtxtext}

\usepackage[T1]{fontenc}

\DeclareRobustCommand{\VAN}[3]{#2}
\let\VANthebibliography\thebibliography
\def\thebibliography{\DeclareRobustCommand{\VAN}[3]{##3}\VANthebibliography}

\def \kms{km s$^{-1}$}
\renewcommand{\ion}[1]{{\normalfont~\textsc{#1}}}
\def \HI{H\ion{i}}
\def \MgEW{${\rm EW_{2796}}$}
\def \FeMg{$\frac{\rm EW_{2600}}{\rm EW_{2796}}$}
\def \nodata{. . .}

\usepackage{lmodern} 
\rmfamily 
\DeclareFontShape{T1}{lmr}{b}{sc}{<->ssub*cmr/bx/sc}{}
\DeclareFontShape{T1}{lmr}{bx}{sc}{<->ssub*cmr/bx/sc}{}

\usepackage{xcolor}

\title{Mapping the spatial extent of \HI-rich absorbers using Mg\ion{ii} absorption along gravitational arcs}
\titlerunning{Mapping the spatial extent of \HI-rich absorbers with Mg\ion{ii}}

\author{
        Trystyn A.~M.~Berg\inst{1}\fnmsep\inst{2}\fnmsep\inst{3}\fnmsep\thanks{Both A.~Afruni and T.~Berg have contributed equally to this paper and are co-first authors. Emails: andrea.afruni@unifi.it, trystyn.berg@nrc-cnrc.gc.ca}
        \and
      Andrea Afruni\inst{2}\fnmsep\inst{4}\fnmsep\inst{5}$^{,\star}$\
      \and
      C\'edric Ledoux\inst{3}
      \and
      Sebastian Lopez\inst{2}
      \and
      Pasquier Noterdaeme\inst{2,6,7}
      \and
      Nicolas Tejos\inst{8}
      \and 
      Joaquin Hernandez\inst{9}
      \and
      Felipe Barrientos\inst{9}
      \and
      Evelyn J. Johnston\inst{10}
      }
\authorrunning{T.~Berg, A.~Afruni et al.}
     \institute{NRC Herzberg Astronomy and Astrophysics Research Centre, 5071 West Saanich Road, Victoria, B.C., Canada, V9E 2E7
     \and
     Departamento de Astronom\'{\i}a, Universidad de Chile, Casilla 36-D, Santiago, Chile.
     \and
     European Southern Observatory, Alonso de Cordova 3107, Casilla 19001, Santiago, Chile.
     \and
     Kapteyn Astronomical Institute, University of Groningen, Landleven 12, 9747 AD Groningen, The Netherlands.
     \and
     Dipartimento di Fisica e Astronomia, Universit\`a di Firenze, Via G. Sansone 1, 50019 Sesto Fiorentino, Firenze, Italy
     \and
     French-Chilean Laboratory for Astronomy, IRL 3386, CNRS
     \and
     Institut d'Astrophysique de Paris, CNRS-SU, UMR,7095, 98bis bd Arago, 75014 Paris, France
     \and
     Instituto de F\'isica, Pontificia Universidad Cat\'olica de Valpara\'iso, Casilla 4059, Valpara\'iso, Chile
    \and
    Instituto de F\'isica, Pontificia Universidad Cat\'olica de Chile, Av. Vicu\~na Mackenna 4860, 7820436 Macul, Santiago, Chile
    \and
    Instituto de Estudios Astrof\'isicos, Facultad de Ingenier\'ia y Ciencias, Universidad Diego Portales, Av. Ej\'ercito Libertador 441, Santiago, Chile
         }

\begin{document}

\abstract{\HI{}-rich absorbers seen within quasar spectra contain the bulk of neutral gas in the Universe. However, the spatial extent of these reservoirs are not extensively studied due to the pencil beam nature of quasar sightlines. Using two giant gravitational arc fields (at redshifts 1.17 and 2.06) as 2D background sources with known strong Mg\ion{ii} absorption observed with the Multi Unit Spectroscopic Explorer integral field spectrograph (IFS), we investigated whether spatially mapped Mg\ion{ii} absorption can predict the presence of strong \HI{} systems, and determine both the physical extent and \HI{} mass of the two absorbing systems. 
We created a simple model of an ensemble of gas clouds in order to simultaneously predict the \HI{} column density and gas covering fraction of \HI{}-rich absorbers based on observations of the Mg\ion{ii} rest-frame equivalent width in IFS spaxels. We first test the model on the lensing field with \HI{} observations already available from the literature, finding that we can recover \HI{} column densities consistent with the previous estimates (although with large uncertainties). We then use our framework to simultaneously predict the gas covering fraction, \HI{} column density and total \HI{} gas mass ($M_{\rm{HI}}$) for both fields.  
We find that both of the observed strong systems have a covering fraction of $\approx70$\% and are likely damped Lyman $\alpha$ systems (DLAs) with $M_{\rm{HI}}>10^9\ M_{\odot}$. Our  model shows that the typical Mg\ion{ii} metrics used in the literature to identify the presence of DLAs are sensitive to the gas covering fraction. However, these Mg\ion{ii} metrics are still sensitive to strong \HI{}, and can be still applied to absorbers towards gravitational arcs or other spatially extended background sources. Based on our results, we speculate that the two strong absorbers are likely representative of a neutral inner circumgalactic medium and are a significant reservoir of fuel for star formation within the host galaxies.}


\maketitle

\section{Introduction}

The gas reservoirs of galaxies are critical for regulating galaxy formation and evolution; from the interstellar medium (ISM) where atomic H\ion{i} cools into molecular gas to form stars \citep[e.g.,][]{saintonge13}, out into the galactic halos where the circumgalactic medium (CGM) hosts both accreted intergalactic gas and ejected material from galactic feedback processes which may continue to feed the ISM \citep[see][and references therein]{Tumlinson17, giguere23}. Thus quantifying the properties of these galactic gas reservoirs across cosmic time is a necessity to constrain our understanding of galaxy evolution.

Quasar sightlines have proved to be a powerful probe of the gas reservoirs of galaxies, enabling a census of the amount of fuel for future star formation and to quantify the metallicity evolution of these reservoirs across cosmic time. Whilst rare in the Universe, the strongest \HI{} absorbers known as damped Ly$\alpha$ systems \citep[DLAs; \HI{} column densities of N(\HI{})~$\geq 2\times10^{20}{\rm cm^{-2}}$;][]{Wolfe05} contain the $\approx80$\% of neutral gas in the Universe \citep{Lanzetta95, Prochaska05, Noterdaeme09, Zafar13, SanchezRamirez16}.

Due to the UV cut-off from Earth's atmosphere, tracing Ly$\alpha$ absorption from galaxies below $z\lesssim1.6$ requires expensive space-based observations. To get access to DLAs in this epoch of the Universe, low-ionization metal species have been used as DLA proxies. The most widely used metal tracer for \HI{}-rich systems is Mg\ion{ii} as it is accessible down to $z\approx0.1$ with ground-based observations \citep{Rao00,Turnshek15}. There are various metrics in the literature for pre-selecting Mg\ion{ii} absorbers at $z\approx1$ as potential DLAs based on the rest-frame equivalent widths (EWs) along quasar sightlines. These metrics include: 

\begin{itemize}
    \item[--] the simple Mg\ion{ii}$\lambda$2796 EW (\MgEW{}). This metric has been used quite extensively in the literature to identify DLAs, as low-redshift DLAs tend to have \MgEW{}$\geq0.3-0.6\AA{}$ \citep{Rao00, Rao06, Turnshek15}. In this work, we focus on using the threshold of  $\geq0.3\AA{}$ which tends to select all DLAs at a variety of redshifts \citep{Matejek13, Berg17, Rao17}.
    
    \item[--] the EW ratio of Fe\ion{ii}$\lambda$2600/Mg\ion{ii}$\lambda$2796 (\FeMg{})\footnote{We note that this ratio is often presented as the inverse -- Mg\ion{ii}$\lambda$2796/Fe\ion{ii}$\lambda$2600 -- in the literature. We opted to use \FeMg\ instead for presentation purposes such that smaller values of the ratio to correspond to lower logN(\HI{}).}. For DLAs, \FeMg{} is found to be between 0.5 and 1.0 \citep{Rao06}. This metric is motivated by the self-shielding nature of DLAs, where the bulk of Fe and Mg will be in the singly-ionized phase. In lower logN(\HI{}) column density absorbers where ionization effects have an effect, the ionizing field will preferentially ionize Fe outside the Fe\ion{ii} state, leading to a smaller \FeMg{} ratio. However, this ratio is also sensitive to the effects of nucleosynthesis (as it traces [$\alpha$/Fe]) and differential dust depletion \citep[e.g.,][]{dey15,Lopez20}. 
    
    \item[--] the $D$-index \citep{Ellison06}. The $D$-index is computed as \MgEW{} normalized by the width of integration bounds for the EW calculation. The minimum $D$-index for which a system is flagged as a potential DLA is set by the spectral resolution used.
\end{itemize}

 While these three metrics are useful in identifying a potential DLA absorber along quasar sightlines, meeting the various cuts on these metrics is no guarantee that the absorption is indeed from a DLA. In \cite{Rao17} and previous works, the \MgEW{} and \FeMg{} are typically very successful at finding DLAs at low redshifts. However, these metrics also identify other strong \HI{} absorbers, such as subDLAs ($19\leq$~logN(\HI{})~$<20.3~{\rm cm^{-2}}$) at high rates, leading to DLA samples with low purity \citep{Rao06,Bouche08}. Furthermore, higher redshift DLAs that are metal-poor or have narrow velocity profiles sometimes do not pass these metrics \citep{Ellison06,Berg17, Rao17}. While the $D$-index metric is designed to take the velocity profile shape into account, the performance of the $D$-index is heavily dependent on the spectral resolution of the instrument, as the success rate of distinguishing a DLA from a sub-DLA decreases with decreasing spectral resolving power \citep{Ellison06}. For example, with the spectral resolving of a high resolution spectrograph (with a line spread function full-width half maximum [FWHM] of $\geq1.4$~\AA), the $D$-index is very efficient at identifying DLAs with $\approx90$\% purity and a 5\% false positive rate \citep[Table 2 in][]{Ellison06}. However, for lower spectral resolutions, such as what is offered with current state-of-the-art integral field spectrograph facilities such as the Multi Unit Spectroscopic Explorer \citep[MUSE][with an average spectral FWHM of 2.7\AA]{Bacon10}, the $D$-index is expected to identify DLAs with a $\approx84$\% success rate and a false-positive rate of $\approx16$\%. Therefore, none of these three metrics on their own can guarantee that a given Mg\ion{ii} absorber is a DLA.

Other than in a few specific cases \citep[][]{Fumagalli15, Neeleman16}, very little is known about the morphology and the distribution of gas in DLAs, as quasar sightlines only provide 1D information of the gas. Whilst rare multiple close project pairs of (lensed) quasars \citep{lopez99,lopez05,Ellison07,Tytler09,Chen14,Krogager18,Rubin18a} or faint extended galaxies as background sources \citep{Bordoloi14,Rubin18b} have been used to assess the physical extent of DLAs or the CGM, these techniques are still limited to 1D skewers of up to a handful of pointings through the same halo. After the advent of sensitive integral field spectrograph like MUSE and KCWI on 8--10m class telescopes, it is now possible to observe faint gravitational arcs as extended background sources to probe the spatial extent of gaseous reservoirs. This so-called gravitational arc tomography has demonstrated the ability to assess how the CGM material is distributed \citep{Lopez18,Lopez20,Mortensen21,Tejos21,FernandezFigueroa22,Afruni23} and in one case it has been used to put constraints on the mass of H\ion{i} gas in a DLA \cite[][further abbreviated as B22]{Bordoloi22}. However, based on the modest wavelength range of current telescope facilities combined with the typical redshifts of gravitational arcs, it is more difficult to study H\ion{i} directly with gravitational arc tomography compared to neutral metal tracers such as Mg\ion{ii}. As for the pencil-beam quasar sightlines, one possibility is to identify possible DLAs using the Mg\ion{ii} metrics described above, but their accuracy has to date never been tested on extended sources.

In this paper, we investigate the use of the three commonly-used Mg\ion{ii} metrics to identify DLAs along extended background sources, aiming to quantify the spatial extent, covering fractions, and \HI{} gas mass of \HI{}-rich absorption. We focus on MUSE observations of two gravitational arcs (one of them being the one analyzed in B22) with strong Mg\ion{ii} absorption. We first verify that these commonly-used Mg\ion{ii} metrics can be used for extended background sources. Next, based on the measured Mg\ion{ii} metrics and using a novel model framework to reproduce the B22 results, we investigate the success of using Mg\ion{ii} metrics to identify DLA absorbers along extended background sources. Based on this modeling, we estimate the covering fraction and the \HI{} gas mass of the two absorbers studied. In Sect.~\ref{observations}, we report the data reduction and processing used to measure equivalent widths; in Sect.~\ref{metricsSec3}, we apply the Mg\ion{ii} metrics to assess the extent of potential DLAs in the two fields; in Sect.~\ref{modelSection}, we describe our model, how we compared the model outputs with the observations and the main results obtained from this comparison; finally, in Sect.~\ref{discussionSection} we outline the limitations and assumptions of our modeling and we discuss the validity of the usual Mg\ion{ii} metrics, while in Sect.~\ref{conclusions} we summarize this work and we report our main conclusions. Throughout this work, we assume a flat $\Lambda$CDM cosmology with $H_{0}=70$ km s$^{-1}$ Mpc$^{-1}$, $\Omega_{\Lambda} = 0.7$ and $\Omega_{M} = 0.3$.

\begin{figure*}
    \centering
    \includegraphics[width=\textwidth]{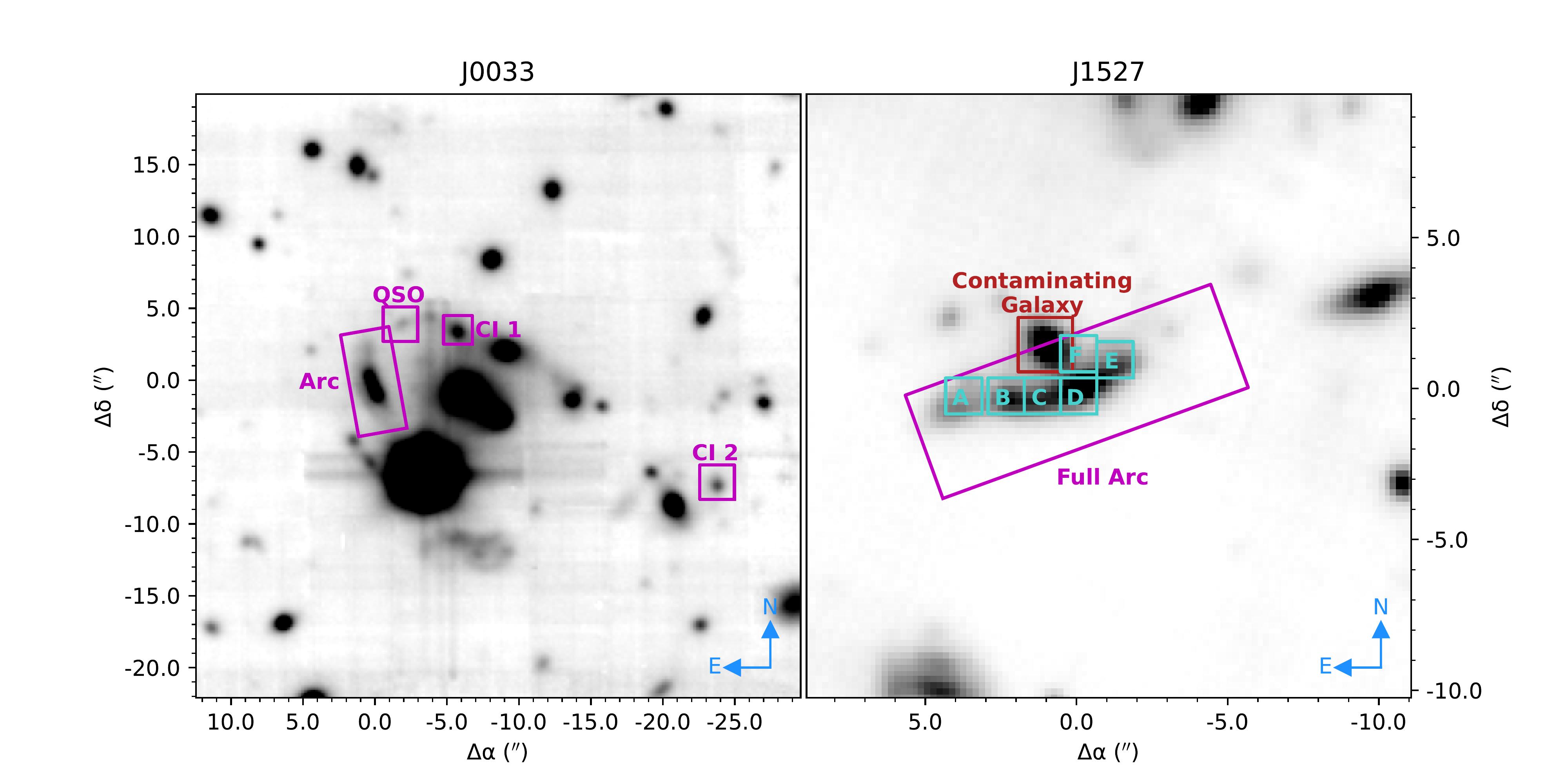}
    \caption{Lensing region towards the J0033 (left panel) and J1527 (right panel) fields. The whitelight image is shown in greyscale. Only a portion of the full MUSE cubes are shown for clarity. The colored magenta rectangles denote regions of spaxels (Table~\ref{tab:regions}) that contain background sources which have detected Mg\ion{ii} absorption. In both panels, the coordinates are relative to the center of the arc region in the respective field (Table~\ref{tab:regions}). For J1527, the smaller cyan boxes denote the six apertures (A--F) from B22 whilst the red rectangle denotes the main region containing a nearby galaxy which contaminates a portion of the gravitational arc. The plate scale of the both images is 0.2\arcsec\ per pixel.}
    \label{fig:fields}
\end{figure*}

\begin{table*}
    \begin{center}
    \caption{Rectangular regions of interest.}
    \label{tab:regions}
    \begin{tabular}{lccccccc}
    \hline
    \hline
    Field & Region & $\alpha$ & $\delta$ & $r_{\alpha}$ & $r_{\delta}$ & Position angle & $\rho_{\perp}$ \\
          &        & ($^{\circ}$)      &($^{\circ}$)   & (``)       & (``)        & ($^{\circ}$) & kpc\\
    \hline
    J0033 & Arc & 8.42321 & 2.70489  & 1.600 & 3.480 &  10.0 & 0.0\\
    J0033 & CI 1 & 8.42158 & 2.70589 &  1.000 & 1.000 & 0.0 & 14.4\\
    J0033 & CI 2 & 8.41658 & 2.70292 &  1.250 & 1.250 & 0.0 & 50.5\\
    J0033 & QSO & 8.42267 & 2.70603 &  1.250 & 1.250 & 0.0 & 12.7\\
    \hline
    J1527 & Full Arc & 231.93796 & 6.87210 &  1.700 & 5.250 &  110.0 & 0.0\\
    J1527 & Galaxy (contaminating Arc) & 231.93825 & 6.87253 &  0.900 & 0.900 & 0.0 & \nodata{}\\
    J1527 & Aperture A & 231.93900 & 6.87204 &  0.587 & 0.587 & 0.0 & 4.7\\
    J1527 & Aperture B & 231.93859 & 6.87204 &  0.587 & 0.587 & 0.0 & 1.1\\
    J1527 & Aperture C & 231.93827 & 6.87204 &  0.587 & 0.587 & 0.0 & 1.1\\
    J1527 & Aperture D & 231.93794 & 6.87204 &  0.587 & 0.587 & 0.0 & 1.4\\
    J1527 & Aperture E & 231.93761 & 6.87237 &  0.587 & 0.587 & 0.0 & 1.8\\
    J1527 & Aperture F & 231.93794 & 6.87245 &  0.587 & 0.587 & 0.0 & 4.6\\
    \hline

    \end{tabular}
    \end{center}
\end{table*}

\section{Observations and data processing}\label{observations}

\subsection{Targets and data reduction}
We observed SGAS J0033+0242  and SGAS J1527+0652 (further referred to as J0033 and J1527, respectively) with the
Multi Unit Spectroscopic Explorer~\citep[MUSE; ][]{Bacon10}) at the Very Large Telescope, as part of programs 098.A-0459 and 0103.A-048, respectively (PI Lopez). These fields were originally selected for having indications of intervening Mg\ion{ii} absorption at $z\sim 1$ on top of the brightest arc knots~\citep{Rigby2018}. The MUSE data revealed strong ($W_0\gtrsim 3 $ \AA) Mg\ion{ii} and Fe\ion{ii} systems at redshifts 1.16729 (J0033) and 2.05601 (J1527), making them good DLA candidates~\citep[e.g.,][]{Rao17}. In both cases the MUSE wide field mode was used, providing a field of view of $1\arcmin \times 1\arcmin$ and a spatial sampling of 0.2\arcsec\,pix$^{-1}$. J0033 was observed without adaptive optics mode and the nominal wavelength range ($\approx 4\,700$--$9\,300$\,\AA), whereas J1527 was observed with the adaptive optics mode and extended wavelength range mode ($4\,600$--$9\,300$ \AA). The on-target exposure times were $2.14$  and $4.17$ hours, respectively. Figure~\ref{fig:fields} shows a zoom-in of the MUSE white-light image of the two fields. Both fields show strong gravitational arcs, whilst the J0033 field contains two lensing counterimages (labelled as CI1 and CI2) as well as a background quasar (QSO).

We reduced the MUSE data using the ESO MUSE pipeline \citep[v2.6,][]{Weilbacher2020} in the ESO Recipe Execution Tool (EsoRex) environment \citep{ESOREX2015}. Besides pre-processing (bias, flat-field, and vacuum-wavelength calibrations), the flux in each exposure was calibrated using standard star observations from the same nights as the science data, and the sky continuum was measured directly from the science exposures and subtracted off. Residual sky contamination was removed from the stacked cube using the Zurich Atmosphere Purge code \citep[][]{Soto2016}. The final cubes were matched to the WCS of the Hubble Space Telescope (HST) images of J0033~\citep{Fischer19} and  J1527~\citep{Sharon20}. These cubes have a point-spread-function of  0.84\arcsec\ and 0.78\arcsec, respectively and a spectral resolving power ranging  from \textit{R}$\simeq$1\,770 at 4\,800\,\AA\ to \textit{R}$\simeq$3\,590 at
9\,300\,\AA.  For further details see \cite{lopez24}.

Table~\ref{tab:regions} contains the coordinates of the relevant regions we select in this work. These regions represent rectangles, centered at the Right Ascension ($\alpha$) and Declination ($\delta$), with a half-width of $r_{RA}$ and $r_{Dec}$ in the respective directions, and rotated by the position angle (measured north through east). Apertures A--F for regions in the J1527 field are the apertures used by B22 (R.~Bordoloi, private communication). The outlines of the spaxels contained by all of these regions are shown as colored rectangles in Fig.~\ref{fig:fields}. In addition, the last column of Table~\ref{tab:regions} provides the delensed projected distance ($\rho_{\perp}$) from the center $\alpha$ and $\delta$ of each region relative to the center $\alpha$ and $\delta$ of the `Arc' region of the respective field. The delensed projected distances are calculated by delensing the image plane to the so called absorber plane (the plane of the absorbing gas), through parametric lens models built for each of the two fields with the software {\sc lenstool}~\citep{Jullo07} and the available HST imaging 
(we refer for more details to \citealt{Sharon20} for J1527 and to \citealt{Fischer19} for J0033).

\subsection{Creating combined spectra from individual spaxels}
\label{sec:combSpec}

For each region of interest within our data (Table~\ref{tab:regions}), we generate flux-weighted combined spectra using all the native spaxels within the region. Spectra for spaxels within the region whose signal-to-noise ratio (S/N, next to spectral region with Mg\ion{ii} absorption) was greater than 1.0 were used to form a combined spectrum. In order to match typical 1D analysis of quasars where the spectral trace is summed together, we first fit and normalize the continuum of the spaxel spectra (see Appendix~\ref{sec:contfit}) and then we weight each spaxel's spectrum by the summed flux of the spectrum at all wavelengths (white-light image\footnote{We opted not to use the \cite{Robertson86} weighting scheme used in previous gravitational arc tomography analyses \citep[e.g.,][]{Lopez18} as the equivalent width extracted for various metal lines were sensitive to the wavelength region selected. We note that the equivalent widths produced using the full wavelength range in the weighting was close to the median equivalent width from selecting different wavelength ranges across the spectrum.}). This weighting is analogous to how pixels are combined along the spatial axis of the slit in 1D quasar sightline observations. We note that each region is larger than the point-spread-function of the observations, and thus the spatial binning from this combination process will minimize the overlapping information from combining the native spaxels. In order to obtain an estimate of the error in the continuum fitting and weighting procedure, we use a Monte Carlo approach by repeating this process 1000 times, where in each iteration the flux of each element of the MUSE cube was sampled from a Gaussian distribution centered on the flux with a standard deviation equal to the $1\sigma$ error spectrum flux. Each of these 1000 spectra are then continuum normalized individually following the method outlined in Appendix \ref{sec:contfit}. We then produce a median spectrum by taking the median flux of each pixel, with the 25$^{\rm th}$ and 75$^{\rm th}$ percentiles to estimate the error.

It is clear from Fig.~\ref{fig:fields} that the regions (particularly the arc of J0033) contain additional background light in and around the selected region after sky subtraction (possibly from extended emission of nearby sources or the nearby bright star, or intra-cluster light). Because of the additional signal, pure-background spaxels can have a S/N~$\geq1$ and thus are selected and included when creating the  combined spectra as described above. By including this background, there is an additional flux within absorption lines that leads to lower EWs and potentially false signatures of emission or partial coverage. In order to remove this additional background prior to making the combined spectra, we first isolate the spaxels (including those with S/N~$<1$) which only contain this additional background flux. This is done by sigma-clipping out all spaxels from the entire region (Table~\ref{tab:regions}) whose white-light image flux is above $+1\sigma$ the mean flux. The spectra of these background-only spaxels are then median combined into a background spectrum that is subtracted from all the spaxels of that region in order to correct for the additional signal. We note that the background signal we observe appears uniformly distributed spatially within the respective regions, and the median combined background spectrum is consistent with these values. Based on this information, we note that the non-uniform light from the galaxy nearby the J1527 arc likely adds additional flux in the spectrum and is not fully subtracted from this method. While we exclude spaxels that contain the galaxy's spectrum  from the analysis (i.e.,~those within the Galaxy region for field J1527 in Table~\ref{tab:regions}), there may still be scattered light issues that add additional flux to the spectrum, particularly within the absorption trough of the key Ly series (KCWI data) and metal lines (MUSE data). 

Figs B.1 -- B.6 in \href{https://zenodo.org/records/14225814}{Appendix B} contain the median metal line profiles from the Monte Carlo analysis for apertures A--F (respectively) in the field J1527 whilst Figs. C.1 -- C.4 in \href{https://zenodo.org/records/14225839}{Appendix C} show the median velocity profiles for the regions within the J0033 field. Both appendices are available online.

\subsection{Measuring rest-frame equivalent widths}
EW measurements are made by integrating the continuum-normalized spectra (see Appendix~\ref{sec:contfit}) between the velocity limits of the absorption feature (Table~\ref{tab:EWvlim}; where the zero-velocity is defined at the redshift of the strongest Mg\ion{ii} absorption, $z_{\rm abs}$). The velocity limits are determined by visually inspecting the absorption of all spaxels and selecting the limits where the strongest line reaches the continuum level. The same velocity limits are adopted for all metal absorption lines of the same system unless there is significant blending in a weak portion absorption profile (e.g., Ca\ion{ii}$\lambda$3934 in Fig. C.1 in \href{https://zenodo.org/records/14225839}{Appendix C}). We adopt the EW derived from the median of the 1000 Monte Carlo spectra, with errors on the EW coming from the the 25$^{\rm th}$ and 75$^{\rm th}$ percentile spectra. 

\begin{table}
    \begin{center}
    \caption{EW velocity integration limits}
    \label{tab:EWvlim}
    \begin{tabular}{lcccc}
    \hline
    \hline
    Field & $z_{\rm abs}$ & Region & $v_{\rm min}$ & $v_{\rm max}$\\
          &  & & (\kms{})     & (\kms{})\\
    \hline
    J0033 & 1.16729& Arc & -425 & 300 \\
    J0033 & 1.16729& CI 1 & -425 & 300 \\
    J0033 & 1.16729& CI 2 & -525 & 250 \\
    J0033 & 1.16729& QSO & -300 & 150 \\
    \hline
    J1527 & 2.05601& Arc & -495 & 400 \\
    J1527 & 2.05601& Aperture A & -400 & 440 \\
    J1527 & 2.05601& Aperture B & -400 & 440 \\
    J1527 & 2.05601& Aperture C & -495 & 440 \\
    J1527 & 2.05601& Aperture D & -495 & 440 \\
    J1527 & 2.05601& Aperture E & -495 & 440 \\
    J1527 & 2.05601& Aperture F & -495 & 440 \\
\hline
    \end{tabular}
    \end{center}
\end{table}

\begin{table*}
\scriptsize
\begin{center}
\caption{Metal line rest-frame equivalent widths for regions of J1527 field}
\label{tab:apertureEWs}
\begin{tabular}{lccccccccc}
\hline
\hline
Region& Mg\ion{i} 2852 
& Mg\ion{ii} 2796 
& Mg\ion{ii} 2803 
& Fe\ion{ii} 2249 
& Fe\ion{ii} 2260 
& Fe\ion{ii} 2344 
& Fe\ion{ii} 2374 
& Fe\ion{ii} 2382 
& Fe\ion{ii} 2600 
\\
& m\AA{}& m\AA{}& m\AA{}& m\AA{}& m\AA{}& m\AA{}& m\AA{}& m\AA{}& m\AA{}\\
\hline
A& $909^{+506}_{-523}$& $3008^{+616}_{-598}$& $2446^{+423}_{-414}$& $<165$& $<248$& $1686^{+244}_{-242}$& $1879^{+380}_{-369}$& $1808^{+483}_{-478}$& $1703^{+423}_{-408}$\\
B& $981^{+307}_{-291}$& $3461^{+361}_{-375}$& $2782^{+259}_{-256}$& $<152$& $479^{+195}_{-196}$& $1386^{+156}_{-164}$& $1507^{+242}_{-243}$& $1892^{+332}_{-319}$& $1946^{+281}_{-269}$\\
C& $978^{+351}_{-349}$& $3272^{+363}_{-356}$& $2740^{+290}_{-281}$& $<179$& $<126$& $1290^{+165}_{-165}$& $1372^{+272}_{-267}$& $2267^{+311}_{-307}$& $2164^{+314}_{-311}$\\
D& $922^{+210}_{-210}$& $3103^{+212}_{-220}$& $2926^{+174}_{-177}$& $<118$& $135^{+136}_{-139}$& $1649^{+114}_{-117}$& $1662^{+181}_{-188}$& $2338^{+202}_{-194}$& $2253^{+203}_{-202}$\\
E& $1005^{+315}_{-318}$& $2546^{+329}_{-340}$& $2646^{+260}_{-253}$& $<208$& $263^{+193}_{-184}$& $1676^{+154}_{-155}$& $1494^{+271}_{-260}$& $1969^{+294}_{-308}$& $2217^{+323}_{-316}$\\
F& $<280$& $1672^{+464}_{-505}$& $1565^{+380}_{-402}$& $<173$& $<214$& $1027^{+256}_{-236}$& $<294$& $2032^{+460}_{-464}$& $1527^{+468}_{-502}$\\
Full arc& $963^{+73}_{-73}$& $2863^{+127}_{-131}$& $2612^{+73}_{-74}$& $<121$& $265^{+75}_{-74}$& $1577^{+54}_{-54}$& $1239^{+103}_{-102}$& $1958^{+117}_{-114}$& $2059^{+103}_{-103}$\\
\hline
\end{tabular}
\end{center}
\end{table*}

\begin{table*}
\scriptsize
\begin{center}

\caption{Metal line rest-frame equivalent widths for regions in J0033 field}
\label{tab:J0033regionEWs}
\begin{tabular}{lcccccccccccc}
\hline
\hline
Region& Mg\ion{i} 2852 
& Mg\ion{ii} 2796 
& Mg\ion{ii} 2803 
& Ca\ion{ii} 3934 
& Ca\ion{ii} 3969 
& Ti\ion{ii} 3384 
& Fe\ion{ii} 2249 
& Fe\ion{ii} 2260 
& Fe\ion{ii} 2344 
& Fe\ion{ii} 2374 
& Fe\ion{ii} 2600 
& Fe\ion{ii} 2382 
\\
& m\AA{}& m\AA{}& m\AA{}& m\AA{}& m\AA{}& m\AA{}& m\AA{}& m\AA{}& m\AA{}& m\AA{}& m\AA{}& m\AA{}\\
\hline
Arc& $955^{+124}_{-120}$& $3668^{+106}_{-111}$& $3592^{+113}_{-112}$& $<1344$& $492^{+160}_{-173}$& $613^{+173}_{-169}$& $<421$& $<180$& $2020^{+107}_{-104}$& $1020^{+99}_{-111}$& $2519^{+112}_{-108}$& $2867^{+94}_{-103}$\\
QSO& $<1229$& $<2732$& $<2161$& $<1536$& $<2967$& $<3535$& $<1819$& $<1653$& $<1143$& $<1027$& $<1244$& \nodata{}\\
CI 1& $2001^{+395}_{-408}$& $3584^{+370}_{-400}$& $3647^{+402}_{-414}$& $<349$& $<352$& $<754$& $<1028$& $<667$& $<3241$& $<1134$& $2825^{+387}_{-379}$& $3038^{+405}_{-404}$\\
CI 2& $<576$& $2206^{+475}_{-498}$& $2504^{+485}_{-494}$& $<420$& $<940$& $<788$& $<667$& $<1299$& $<1254$& $<623$& $2106^{+436}_{-440}$& $1853^{+397}_{-402}$\\
\hline

\end{tabular}
\end{center}
\end{table*}

We require that the rest-frame equivalent widths be measured at $>2\sigma$ (based on the S/N of the median spectrum) to be considered as detected. Otherwise, $2\sigma$ upper limits are adopted. Lines that are strongly blended use the measured equivalent width, but are flagged as upper limits. The median equivalent width measurements for the B22 apertures and full arc in the J1527 field as well as the J0033 regions are provided in Tables \ref{tab:apertureEWs} and \ref{tab:J0033regionEWs}, respectively. The same equivalent width information is also provided online in \href{https://zenodo.org/records/14225814}{Appendix B} (J1527) and \href{https://zenodo.org/records/14225839}{Appendix C} (J0033).

\section{DLA identification using Mg\ion{ii} metrics}\label{metricsSec3}

\subsection{Verifying Mg\ion{ii} DLA metrics on extended sources within the J1527 field}

The Mg\ion{ii} DLA identification metrics (i.e.,~\MgEW{}, \FeMg{}, and $D$-index; defined in the introduction) have been defined on quasar sightlines. It remains untested whether these three metrics can also be used for identifying DLAs in multiplexed data of extended sources such as gravitational arcs. We take advantage of testing these metrics towards the J1527 field, where existing KCWI observations have demonstrated that all six apertures (apertures A--F; Table~\ref{tab:regions}) host a DLA whilst our MUSE data complement the KCWI data by providing coverage of several key metal lines, including the Mg\ion{ii} doublet.  Using these six aperture regions,  we created combined spectra following the method outlined in Sect.~\ref{sec:combSpec}.  The EWs of the key metal lines are tabulated in Table~\ref{tab:apertureEWs} while the resulting DLA metrics are provided in Table~\ref{tab:DLAmetricJ1527} along with the \HI{} column densities from B22.

\begin{table}
\begin{center}
\caption{DLA metrics for B22 apertures of J1527 field}
\label{tab:DLAmetricJ1527}
\begin{tabular}{lcccc}
\hline
\hline
{\small Aperture}& logN(\HI{})$^{\rm a}$& ${\rm EW_{2796}}$& $\frac{\rm EW_{2600}}{\rm EW_{2796}}$& $D$\\
& [cm$^{-2}$]& \AA{}& & \\
\hline
A& $20.90\pm0.30$& ${\bf 3.01^{+0.62}_{-0.60}}$& ${\bf0.57^{+0.32}_{-0.21}}$& ${\bf3.8^{+0.8}_{-0.7}}$\\
B& $20.90\pm0.30$& ${\bf3.46^{+0.36}_{-0.38}}$& ${\bf0.56^{+0.16}_{-0.12}}$& ${\bf4.1^{+0.4}_{-0.4}}$\\
C& $20.75\pm0.30$& ${\bf3.27^{+0.36}_{-0.36}}$& ${\bf0.66^{+0.19}_{-0.15}}$& $3.5^{+0.4}_{-0.4}$\\
D& $20.80\pm0.30$& ${\bf3.10^{+0.21}_{-0.22}}$& ${\bf0.73^{+0.13}_{-0.11}}$& $3.3^{+0.2}_{-0.2}$\\
E& $20.90\pm0.30$& ${\bf2.55^{+0.33}_{-0.34}}$& ${\bf0.87^{+0.28}_{-0.21}}$& $3.0^{+0.4}_{-0.4}$\\
F& $20.80\pm0.30$& ${\bf1.67^{+0.46}_{-0.51}}$& ${\bf0.91^{+0.80}_{-0.43}}$& $2.0^{+0.5}_{-0.6}$\\
Full arc& \nodata\ & ${\bf2.86^{+0.13}_{-0.13}}$& ${\bf0.72^{+0.07}_{-0.07}}$& $3.2^{+0.1}_{-0.1}$\\
\hline
\end{tabular}
\end{center}
Notes: $^{\rm a}$logN(\HI{}) taken from B22.
\end{table}

\begin{table}
\begin{center}
\caption{DLA metrics for regions in J0033 field}
\label{tab:J0033DLAmetric}
\begin{tabular}{lccc}
\hline
\hline
Region& ${\rm EW_{2796}}$& \FeMg{} & $D$-index\\
& \AA{}& & \\
\hline
Arc& ${\bf3.67^{+0.11}_{-0.11}}$& ${\bf0.69^{+0.05}_{-0.05}}$& ${\bf5.1^{+0.1}_{-0.2}}$\\
CI 1& ${\bf3.58^{+0.37}_{-0.40}}$& ${\bf0.79^{+0.22}_{-0.17}}$& ${\bf4.9^{+0.5}_{-0.6}}$\\
CI 2& ${\bf2.21^{+0.48}_{-0.50}}$& ${\bf0.95^{+0.53}_{-0.33}}$& $2.8^{+0.6}_{-0.6}$\\
QSO& $<2.73$& \nodata{}& $<6.1$\\
\hline
\end{tabular}
\end{center}
\end{table}

 For all six B22 apertures (which all host a DLA based on the logN\HI{} measured in B22) and the full arc region, the \MgEW{} and \FeMg{} metrics tabulated in Table \ref{tab:DLAmetricJ1527} are consistent with the apertures containing a DLA using the standard quasar metric thresholds, suggesting that these two metrics from \cite{Rao00} can be used to identify DLAs. Based on the MUSE instrumental FWHM (2.54~\AA{}) at the wavelength of Mg\ion{ii} 2796\AA{}, the expected $D$-index threshold for selecting a DLA is $\approx3.6$ \citep[interpolating Table 2 of][]{Ellison06}. This $D$-index threshold would only select apertures A--C as hosting DLAs. However, we caution that at this low spectral resolution, the success rate of identifying DLAs is expected to be $\leq84$ per cent \citep{Ellison06}; the observed success rate of the $D$-index for these six apertures is 33\%. All entries in Table~\ref{tab:DLAmetricJ1527} that would be flagged as a DLA based on the respective Mg\ion{ii} metric have been bolded. Combining all the spaxels together from full arc in place of the individual six B22 apertures, we obtain similar results -- both the \MgEW{} and \FeMg{} metrics are consistent with hosting DLAs, whilst the $D$-index ($3.2\pm0.1$) is just below the threshold for selecting a DLA. We point out that the value of all three metrics for the combined full arc spectrum are effectively equal to the average of the values derived for the six B22 apertures.

As mentioned in Sect.~\ref{sec:combSpec}, there is potential contamination in all apertures from galaxy light close in projection to the arc. As we are using weighting from the white-light image, the spaxels that contain more galaxy light than arc light are favored, and artificially weaken the Mg\ion{ii} absorption in the combined spectrum by adding flux within the absorption trough. We therefore caution that EW measurements in the apertures (Table~\ref{tab:apertureEWs}; particularly aperture F) are likely lower limits to the true value arising from pure Mg\ion{ii} absorption. This may partially explain why the $D$-index is just below the threshold for selecting a DLA and is not as successful at identifying the DLA absorption in this particular field. It is possible that this contamination issue also affects the determination of the \HI{} column density and the presence of partial coverage reported by B22.

\begin{figure*}
    \centering
    \includegraphics[width=\textwidth]{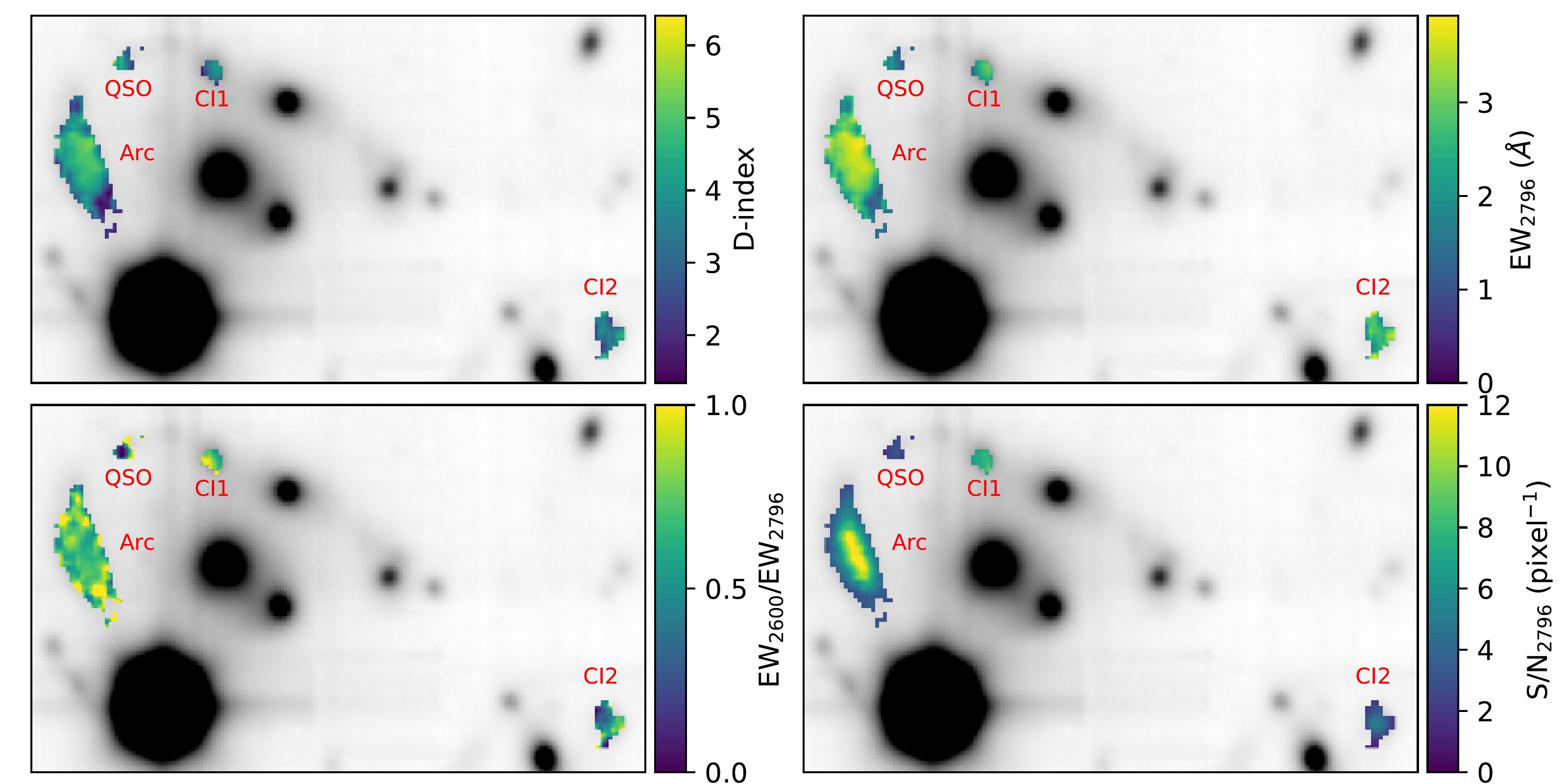}
    \caption{Color maps of the lensed region in the J0033 field, showing the four different regions (Arc, QSO, CI1, and CI2) tracing Mg\ion{ii} absorption.  The top left ($D$-index), top right (Mg\ion{ii}$\lambda$2796 rest-frame equivalent width), and bottom left (Fe\ion{ii}$\lambda$2600/Mg\ion{ii}$\lambda$2796 equivalent width ratio) panels show three different metrics used in the literature to identify if the detected Mg\ion{ii} absorption is a potential DLA. The color bars are designed such that spaxels that are green and yellow colors in these three panels would be considered as potential DLA absorption. The bottom right panel shows the S/N at the position of the Mg\ion{ii} 2796 \AA{} absorption at each spaxel. In all four panels, the white-light image of the cube is shown for reference. Only spaxels with S/N~$\geq2.0$ at the position of Mg\ion{ii} 2796 \AA{} and \MgEW{} detected at $\geq2\sigma$ significance are colored in all four panels. }
    \label{fig:DLAmapJ0033}
\end{figure*}

\begin{SCfigure*}[0.5]
    \centering
    \includegraphics[width=0.85\textwidth]{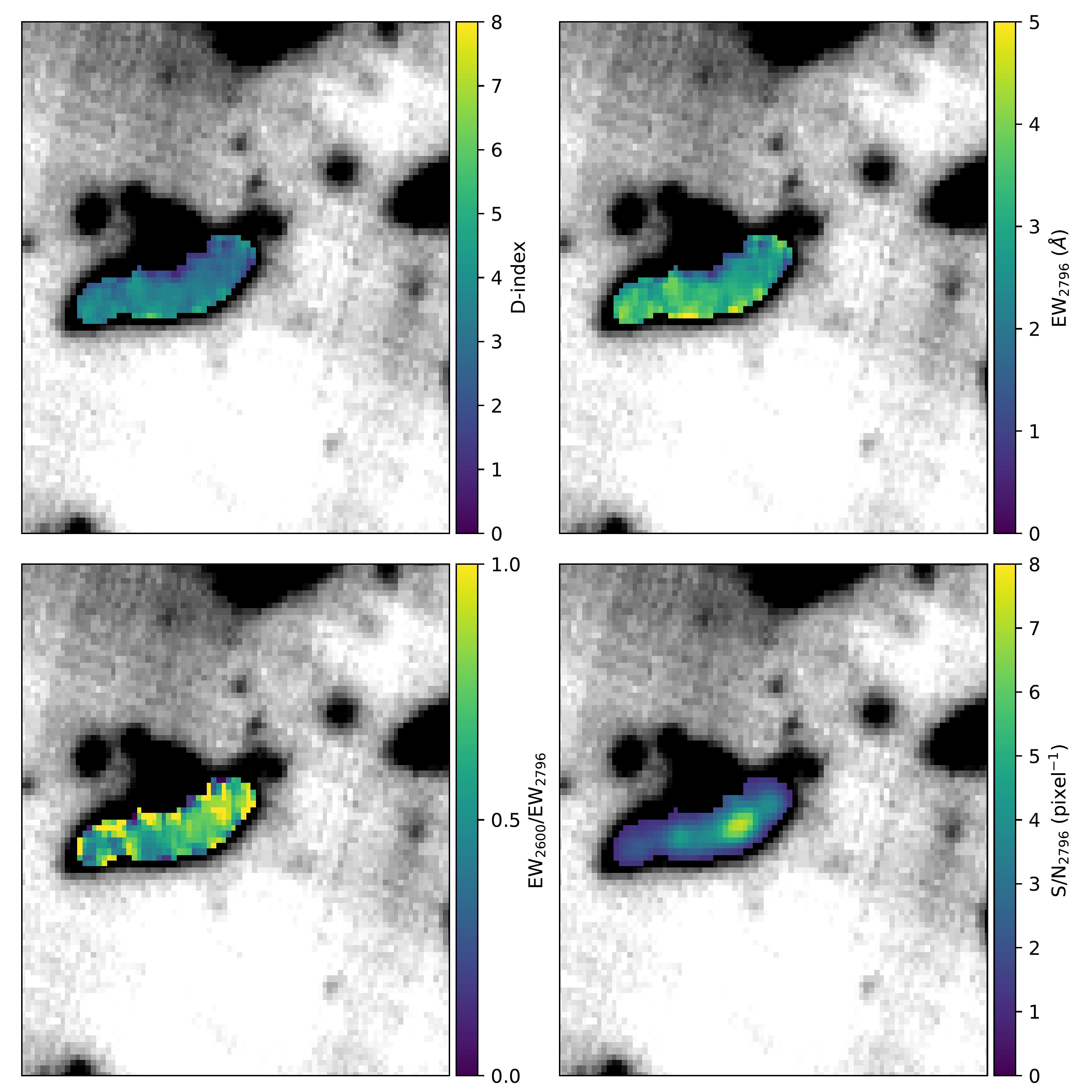}
    \caption{Color maps of the lensed region in the J1527 field, showing the gravitational arc tracing Mg\ion{ii} absorption.  The notation is the same as in Fig.~\ref{fig:DLAmapJ0033}. Only spaxels with S/N~$\geq1$ at the position of Mg\ion{ii} 2796 \AA{} and a \MgEW{} detected at $\geq2\sigma$ significance are colored in all four panels. }
    \label{fig:DLAmapJ1527}
\end{SCfigure*}

\subsection{Applying Mg\ion{ii} metrics to the J0033 field}

 Following the same methodology for the J1527 field, we created combined spaxel spectra for all four regions and measured the EWs for the same metal lines (Table~\ref{tab:J0033regionEWs}) and measured the Mg\ion{ii} metrics for DLA identification (Table~\ref{tab:J0033DLAmetric}). Bolded entries in Table~\ref{tab:J0033DLAmetric} represent regions where the given Mg\ion{ii} metric would be flagged as a DLA. In the combined spectra from three of the four background sources of the lensing system towards J0033 (i.e.,~the Arc, CI 1, and CI 2 regions in Table~\ref{tab:regions}), there is very strong  Mg\ion{ii}$\lambda$2796 absorption (rest-frame equivalent widths $>2\ \AA{}$; Figs. C.1 -- C.3 in \href{https://zenodo.org/records/14225839}{Appendix C}) that suggests the presence of an \HI{}-rich absorber based on the results from J1527. Unfortunately the combined spectrum for the QSO region has poor S/N (Fig. C.4 in \href{https://zenodo.org/records/14225839}{Appendix C}), so we exclude this background source from our analysis. What makes this particular field interesting is that large physical separations are probed by the arc and two counter images (separated by $\approx14$--50 kpc), compared to the $\approx10$~kpc scales measured across the J1527 arc (see Table~\ref{tab:regions}). Despite not having access to the Lyman series to measure the H\ion{i} column density directly, the J0033 field provides an opportunity to study the spatial extent of a strong Mg\ion{ii} absorber that is likely a DLA (or sub-DLA) based on the very strong Mg\ion{ii} absorption (\MgEW{}~$\gtrsim2$~\AA{}).

In Figs. C.1 -- C.3 in \href{https://zenodo.org/records/14225839}{Appendix C}, we note there is no obvious difference in shape or centroid velocity of the Mg\ion{ii}$\lambda$2796 absorption profile between the four regions. The kinematics of this particular absorber will be discussed in more detail in a future paper (Ledoux et al., in prep.). While the spectral resolution of MUSE is insufficient to determine if sub-components of the velocity profile are in agreement between all three regions, the centroids are consistent within $\approx 50$~\kms{} of each other (i.e.,~well within the spectral FWHM of MUSE), and (apart from the QSO region, which has low S/N) all regions appear to show a velocity profile with potentially at least two components. This would suggest that the stronger absorber has the same bulk kinematics across the entire $\approx50$ kpc extent traced by the background sources (Table~\ref{tab:regions}).

\subsection{Mapping Mg\ion{ii} DLA metrics with MUSE spaxels}\label{Musemaps}

In order to study the physical extent of the two strong Mg\ion{ii} absorbers, Figs.~\ref{fig:DLAmapJ0033} and \ref{fig:DLAmapJ1527} show spaxels maps of each of these three metrics in the fields of J0033 and J1527 respectively. The spectra generated for each pixel (corresponding to a native spaxel) in these maps are created using the same method outlined in Sect.~\ref{sec:combSpec}, and are combined using nine spaxels within a $3\times3$ spaxel grid centered at each native pixel. This procedure effectively corresponds to a 0.6\arcsec $\times$ 0.6\arcsec smoothing of the original map, averaging over spaxels contained within the point spread function of the observations ($\approx0.8$\arcsec). This smoothing thus minimizes effects from overlapping information within the native pixel scale. Despite this smoothing, we point out that there are only $\approx30$ and $\approx44$ spatially-independent measurements within the respective J0033 and J1527 fields, which we discuss in more detail in Sect.~\ref{limitations}.

We note that there is a potential correlation between each of the three metrics and the measured S/N near the Mg\ion{ii} doublet for spaxels with S/N~$<1$ and an EW detection significance $<2\sigma$ in both fields. We thus proceed by implementing a quality control cut, and only analyze spaxels with a S/N~$\geq1$ near the Mg\ion{ii} doublet and a \MgEW{} detected at $\geq2\sigma$ significance.

\begin{table*}[]
    \centering
    \caption{Percentage of spaxels that pass the Mg\ion{ii} metric in each region}
    \label{tab:spaxstats}
    \begin{tabular}{lcccc}
    \hline
    \hline
    Metric(s) & J0033 arc & J0033 CI1 & J0033 CI2& J1527 full arc \\
     & \% & \% & \% & \% \\
    \hline
    \MgEW\  &  100 & 100 & 100 & 100\\
    \MgEW\ $+$ \FeMg\ & 76& 89& 44&66\\
    $D$-index & 57& 36& 35& 33\\
    \MgEW\ $+$ \FeMg\ $+$ $D$-index & 34& 26& 12&16\\
    \hline
    \end{tabular}
\end{table*}

Table \ref{tab:spaxstats} provides the percentage of spaxels in Figs.~\ref{fig:DLAmapJ0033} and \ref{fig:DLAmapJ1527}  that pass the various Mg\ion{ii} metric criterion typically used to flag an absorber as a potential DLA. In each of the three regions of J0033 and the full arc of J1527, 100\% of spaxels that meet our quality control cut pass the \MgEW{}~$ \geq 0.3$~\AA{}. This would suggest that this candidate DLA is completely extended over the background source. However combining with other metrics, (i) 44--89\% of spaxels passed both the \MgEW{} and \FeMg\ criteria for potential DLA absorption, and (ii) 12\%–34\% of spaxels simultaneously meet the $D$-index, \MgEW{} and \FeMg\ criteria for potential DLA absorption. We note that, of the 12\% of spaxels that pass all three metric criteria within the CI2 region of the J0033 field, all spaxels are found on the outer edge of the region where the S/N is lowest (S/N~$\approx2$). However, most of the spaxels that pass the three Mg\ion{ii} metric criteria for the arc in both fields and CI1 region in J0033 are more evenly distributed across the region. Assuming a combination of all three of the Mg\ion{ii} metrics is a definitive tracer for \HI{}-rich absorption, the lower limit on the overall extent of \HI{} gas across all three regions of the J0033 field is $\gtrsim30$\%, and $\gtrsim 16$\% for the full arc towards J1527 \citep[prior to accounting for the predicted false positive and success rates of each of the three metrics; e.g.,~as seen for the $D$-index in][]{Ellison06}.

\section{Modeling the fields}\label{modelSection}
While in the previous sections we analyzed the direct observational properties of the strong Mg\ion{ii} absorbers detected in the J1527 and J0033 fields, in the following we use a novel approach that makes use of Bayesian inference and parametric physical models to interpret the detected strong absorption. The main goals of this analysis are to i) infer the physical properties of the absorbers and ii) test the robustness of Mg\ion{ii} metrics and whether they can be applicable on extended sources (see Sect.~\ref{DiscussionMetrics}).
 
In our modeling procedure we assume that the two Mg\ion{ii} absorbing systems are physically represented by a population of cool clouds, from which we extract mock observations (specifically EWs and $D$-indices that one would extract from an ensemble of spaxels) that can be directly compared to the real data.  
These models are idealized and, while they are useful to make conclusions on the two dimensional properties of the absorbing gas (e.g.,~covering fraction), they do not have the pretense of depicting a realistic three dimensional cloud distribution. In Sect.~\ref{limitations} we discuss more in detail the various assumptions and limitations of the models.

\begin{figure}
    \centering
    \includegraphics[clip, trim={5cm 0.5cm 45cm 1cm}, width=\columnwidth]{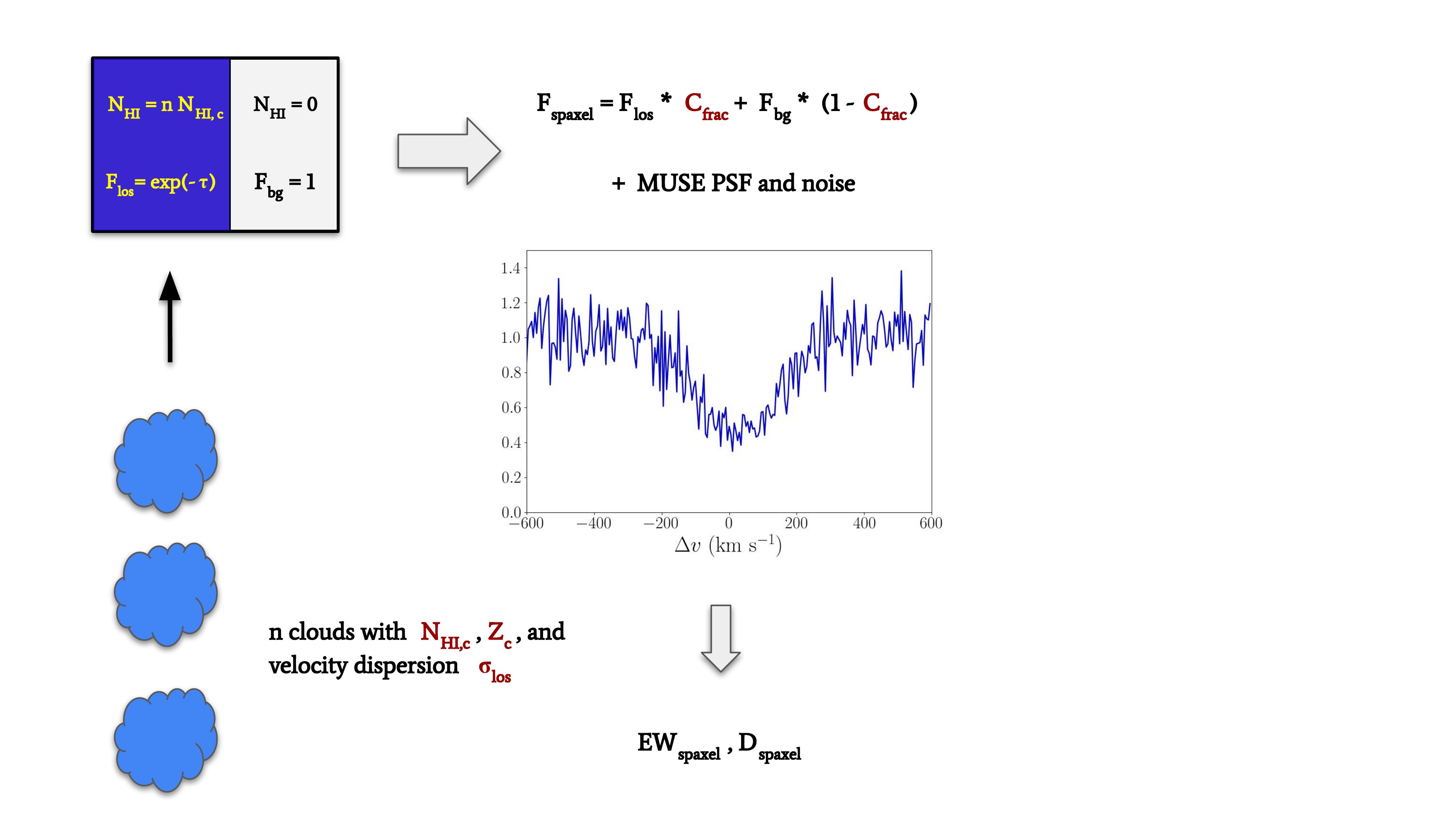}
    \caption{Schematics showing how we extract a pair of EW and $D$-index from a single spaxel, given a choice of the free parameters (highlighted in red) of our modeling. An ensemble of $n$ clouds produces the absorption. The spaxel is only partially covered by gas, so that the flux is either equal to $F_{\rm{los}}=F_{\rm{bg}}\exp{(-\tau)}$ if the clouds are intercepted or to $F_{\rm{bg}}=1$ if they are not. The total flux of the spaxel $F_{\rm{spaxel}}$ depends on the value of the covering fraction $C_{\rm{frac}}$.
    For more details, see Sect.~\ref{procSP}}
    \label{fig:sketchModels}
\end{figure} 

\subsection{Model description}
\label{sec:model}
\subsubsection{Procedure for a single spaxel}\label{procSP}
We first assume that the flux detected by a spaxel is directly related to the gas covering fraction $C_{\rm{frac}}$ within that spaxel (hence the fraction of the background source area covered by the gas) through the formula:
\begin{ceqn}
\begin{equation}\label{eq:flux}
F_{\rm{spaxel}}=C_{\rm{frac}}F_{\rm{los}}+F_{\rm{bg}}(1-C_{\rm{frac}})\  .
\end{equation}
\end{ceqn}
Here, $F_{\rm{bg}}$ is the background continuum flux, which we assume to be normalized to 1, while $F_{\rm{los}}$ is the flux that would be detected by a single point-source line of sight that intercepts the absorbing gas. For Eq.~\eqref{eq:flux} to be valid, we assume that the background flux is constant within the spaxel and that the absorbing gas, where present, has exactly the same properties within the same spaxel. This is clearly an approximation, which however should not significantly impact the general results of this paper. We emphasize that, in this paper, we define the covering fraction as the fraction of the area of the background source area (such as a spaxel) that is covered by foreground gas. This is different to a more observationally-driven definition with the fraction of spaxels across an extended source that contain the same absorption feature, such as a DLA. The MUSE spaxels have a size of (0.2\arcsec $\times$ 0.2\arcsec), resulting in typical physical sizes in both fields of 0.07 and 1.4 kpc along a side (after delensing). The output flux of the model is anyway independent of the spaxel size itself and is only affected by the covering fraction $C_{\rm{frac}}$ (see Eq.~\ref{eq:flux}).

Once a covering fraction is assumed, to calculate $F_{\rm{spaxel}}$ one needs to calculate the value of $F_{\rm{los}}=F_{\rm{bg}}\exp{(-\tau)}$, where the optical depth $\tau$ is defined as \citep[see e.g.,][]{liang17}:
\begin{ceqn}
\begin{equation}\label{eq:optdepth}
\tau(\lambda|N,b,v) = N \sigma_0 f_{\rm{osc}} \Phi(\lambda|b,v)\ ,
\end{equation}
\end{ceqn}
where $N$ is the gas column density, $\sigma_0$ is the cross section, $f_{\rm{osc}}$ is the oscillator strength and $\Phi(\lambda|b,v)$ is the Voigt profile function, which depends on the wavelength $\lambda$, on the Doppler parameter $b$ and on the velocity $v$ at which the transition takes place.  
We assume that the absorption is due to the superposition of $n$ clouds along the line of sight, each of them contributing to the total optical depth. We first assume that each of these clouds has a neutral hydrogen column density $N_{\rm{HI,c}}$, so that the total HI column density along the point-source line of sight would be $n N_{\rm{HI,c}}$. We then define the spaxel HI column density $N_{\rm{HI,spaxel}}$ as $n N_{\rm{HI,c}} \times C_{\rm{frac}}$, given that the spaxel is only partially covered by gas. Assuming a value for $N_{\rm{HI,spaxel}}$, one can then solve for the number of clouds:
\begin{ceqn}
\begin{equation}\label{eq:covfrac}
    n = \frac{N_{\rm{HI,spaxel}}}{C_{\rm{frac}} N_{\rm{HI,c}}}\ .
\end{equation}
\end{ceqn}
The goal of this modeling is to find which HI column densities and covering fractions are needed to reproduce the observed Mg\ion{ii} EWs and $D$-indices. To this purpose, we further assume that the clouds have a metallicity $Z_{\rm{c}}$ and an intrinsic volumetric density $n_{\rm{c}}$ \footnote{Throughout this work, we assume for simplicity that all the clouds have $n_{\rm{c}}=0.01$ cm$^{-3}$. Changing this value (which is essentially unknown from observations) does not significantly impact our results, in particular the values of the covering fraction. As we discuss in Sect.~\ref{ModelResults}, the recovered HI column densities and cloud metallicities are loosely constrained. Hence, even though the photo-ionization models will give different outputs for different values of the intrinsic density, our overall conclusions are not affected by this choice.}. We then use the photo-ionization code CLOUDY \citep{ferland13} to infer the Mg\ion{ii} column densities of the single clouds, assuming an ionizing flux given by the extragalactic UV background from \cite{haardt12} at the redshift of the absorbers (see Sect.~\ref{observations}). At this high column densities most of the medium will be self-shielded and not affected by ionization. Models that consider completely neutral gas lead us to the same results and conclusions presented here (we discuss this in more detail in Sect.~\ref{limitations}). We then assign to each of the $n$ clouds a line-of-sight velocity extracted from a Gaussian distribution centered in zero and with a width equal to a velocity dispersion $\sigma_{\rm{los}}$ and a Doppler parameter $b$ equal to the sum in quadrature of a thermal ($b_{\rm{th}}\sim4$ \kms{}, assuming a temperature $T=2\times10^4$ K) and a turbulent component, with $b_{\rm{turb}}=10$ \kms{} (we discuss different values of $b_{\rm{turb}}$ in Sect.~\ref{limitations}). As we discuss below, $\sigma_{\rm{los}}$ is one of the free parameters of our model and it determines the (simplistic) bulk kinematics of the cloud population and, in turn, the strength of the absorption. With all the above ingredients, one can solve Eqs.~\eqref{eq:optdepth} and \eqref{eq:flux} to finally obtain the spaxel flux $F_{\rm{spaxel}}$, specifically for the Mg\ion{ii} absorption lines.

As a final step, we convolve the spectrum with a Gaussian kernel assuming that the instrumental profile has a resolution (FWHM) of 150 \kms{}, which roughly resembles the resolution of MUSE at the redshift of the two absorbers. We then add random Gaussian noise in the spectrum: the value of the final flux at each wavelength is extracted from a Gaussian centered on the original model flux value, with a width equal to the flux value divided by the observed signal-to-noise in the MUSE data. Once this procedure is complete, we can extract the Mg\ion{ii} metrics of the spaxel, specifically the \MgEW{} and the $D$-index. We point out that we do not include the ratio \FeMg\ in this part of the analysis: this ratio might be dependent on our assumption of photo-ionization (which we discuss in Sect.~\ref{limitations}) and is additionally also affected by dust and by chemical enrichment, which are not explicitly taken into account in our framework. The steps explained above are summarized in the schematics of Fig.~\ref{fig:sketchModels}.

\begin{figure*}
    \centering    \includegraphics[width=0.49\textwidth]{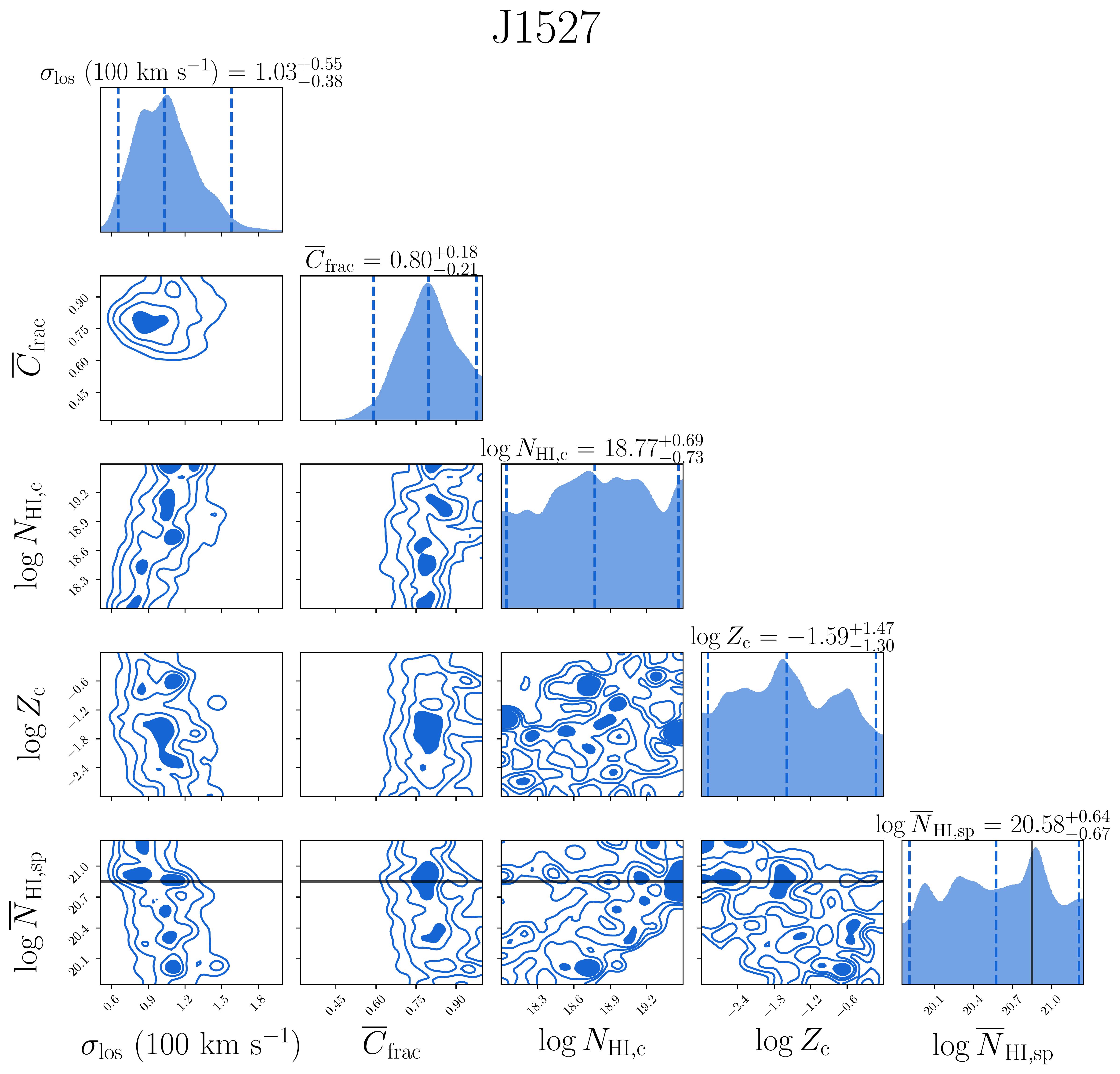}
\includegraphics[width=0.49\textwidth]{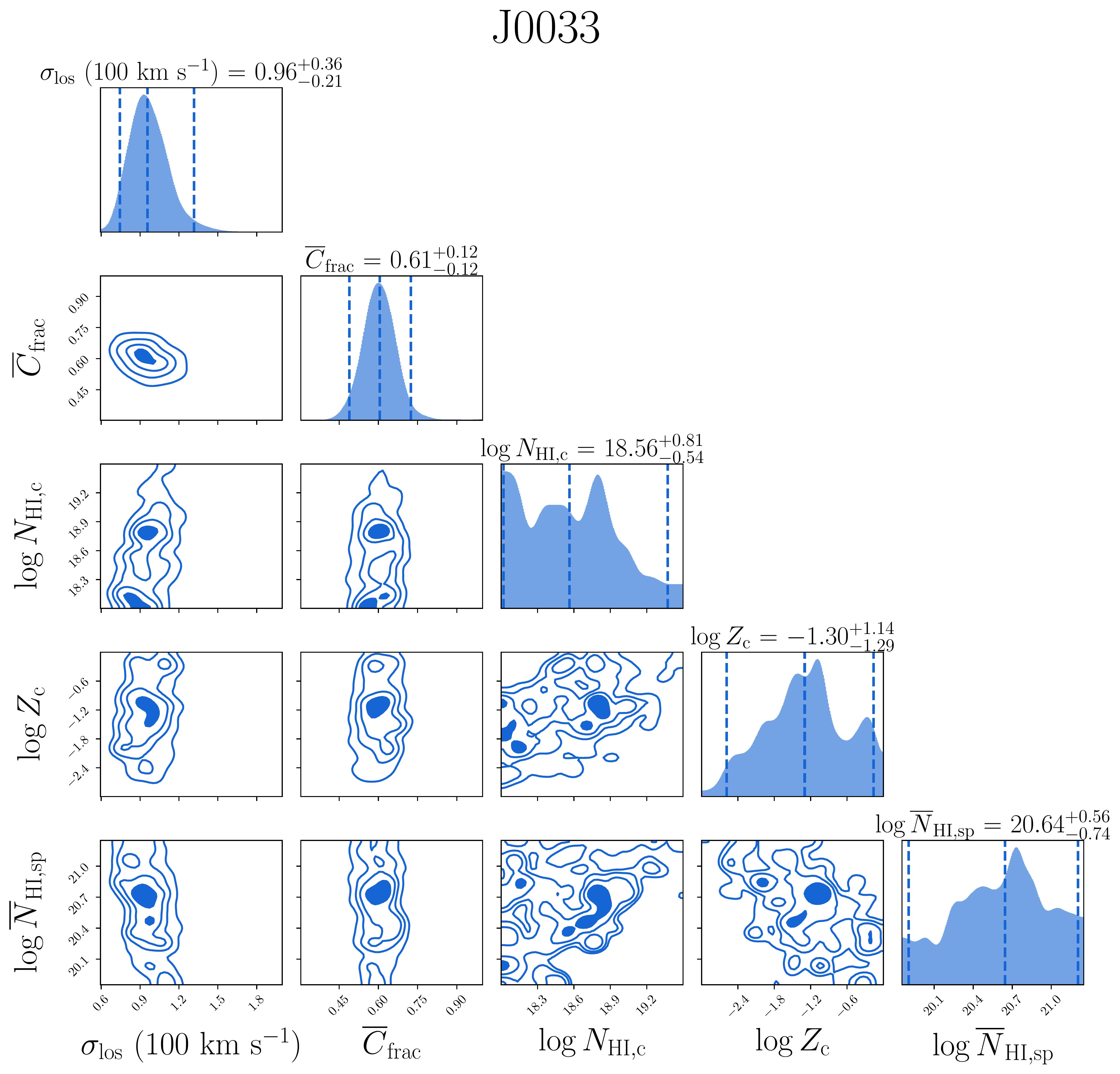}
    \caption{Posterior distributions resulting from the Bayesian analysis described in Sect.~\ref{ModelResults}. 
    The blue vertical dashed lines show the median and 2$\sigma$ uncertainties of each distribution and the black line marks the HI column density estimate of \cite{Bordoloi22}, which is consistent with our predictions.}\label{fig:CornerPlotsJ1527}
\end{figure*}

\subsubsection{Prediction from an ensemble of spaxels and comparison with the data}

To compare with the observations outlined in the previous sections, we need to apply the procedure above to multiple spaxels. The Mg\ion{ii} metrics extracted from this spaxel ensemble are then directly comparable to our data. In the following, we assume that for each model realization and across the same field all the clouds of our synthetic populations have the same metallicities and densities, with velocities extracted from the same Gaussian velocity distribution. However, in Sect.~\ref{Musemaps} we have seen that the observed properties of the fields vary across them and that for example a fraction of the spaxels do not pass the three metrics for potential DLA absorption. We therefore assume that the differences across the spaxels are due to differences in the spaxel covering fraction and in the total HI column density (i.e., different spaxels detect different numbers of clouds). We introduce the parameters
$\overline{N}_{\rm{HI,spaxel}}$, which is the mean observed neutral hydrogen column density of the spaxels (the column density of each spaxel is then drawn randomly from a range that goes from $\log (\overline{N}_{\rm{HI,spaxel}}) - 0.3$ to $\log (\overline{N}_{\rm{HI,spaxel}}) + 0.3$) and $\overline{C}_{\rm{frac}}$, which is the center of a normal distribution with a standard deviation equal to 0.2 and truncated at 0.1 and 1, from which the value of the covering fraction $C_{\rm{frac}}$ of each spaxel is extracted.

A single model realization, hence a distribution of \MgEW{} and $D$-indices (see Sect.~\ref{procSP}), is uniquely defined by the choice of the five free parameters $\overline{N}_{\rm{HI,spaxel}}$, $\overline{C}_{\rm{frac}}$, $Z_{\rm{c}}$, $N_{\rm{HI,c}}$ and $\sigma_{\rm{los}}$. Each distribution is composed of 200 values, a number consistent with the amount of spaxels in J1527 and J0033 (see Section~\ref{Musemaps}). 
The idea is then to compare these distributions with the observed ones to find which choice of parameters better reproduce our data. In order to do this, we first select the spaxels in the data and in the model realizations using the same quality control cuts as used for the observations (i.e., S/N~$>1$ and an \MgEW{} detection significance $>2\sigma$\footnote{For the model, we calculate this uncertainty for each spaxel by bootstrapping the noise into the model flux and by taking the standard deviation of the resultant distribution of EWs.}). We then quantitatively assess the consistency between the model predictions and the MUSE observations by performing a Kolmogorov-Smirnov (KS) test on the distributions of EWs and $D$-indices.  Finally, we define our likelihood as:

\begin{ceqn}
\begin{equation}\label{eq:likelihood}
\ln{\mathcal{L}}  = \ln{p_{EW}} + \ln{p_{D}}\ ,
\end{equation}
\end{ceqn}

\noindent where $p_{EW}$ and $p_{D}$ are the p-values\footnote{We opted to use the p-value in place of the KS test metric to ensure the likelihood is more sensitive to small changes in the KS test metric when the distributions are most similar. We also acknowledge that Eq.~\eqref{eq:likelihood} assumes that EWs and $D$-indices are independent, but these two quantities depend on each other, given the definition of the $D$-index.} obtained from the KS tests on respectively the EW and $D$-index distributions. We finally use the likelihood expressed in Eq.~\eqref{eq:likelihood} to perform two Bayesian analyses on the J1527 and the J0033 fields, whose results are presented in the next section. We point out that the results of our analysis (i.e., the recovered best-fit values of the five free parameters) are independent from any lensing model, as we are not assuming a priori the size and shape of the two fields or of the single spaxels.

\subsection{Model results}\label{ModelResults}
In this section, we present the results of the Bayesian analysis performed to compare our models with our two observational fields, as explained in the previous section. We adopt the nested sampling method \citep{skilling04,skilling06}, using the {\sc dynesty} python package \citep[][]{speagle20,koposov22}. For both fields, we use flat priors for the five free parameters: $19.8<\log \overline{N}_{\rm{HI,spaxel}}<21.2$, $50<\sigma_{\rm{los}}/(\rm{km}\ \rm{s}^{-1})<200$, $0.1<\overline{C}_{\rm{frac}}<1$, $18<\log N_{\rm{HI,c}}<19.5~{\rm cm^{-2}}$, $-3<\log Z_{\rm{c}}<0$.

\subsubsection{J1527}
The results of the Bayesian analysis performed on the J1527 field are shown in the left-hand side of Fig.~\ref{fig:CornerPlotsJ1527}. We can first look at the intrinsic properties of the gas clouds, $\sigma_{\rm{los}}$, $N_{\rm{HI,c}}$ and $Z_{\rm{c}}$. As expected, these three quantities are slightly degenerate with each other, as they all contribute to the strength of the absorption (Eq.~\ref{eq:optdepth}).  At fixed $N_{\rm{HI,spaxel}}$ a lower value of $N_{\rm{HI,c}}$ simply implies a larger number of clouds along the line of sight (Eq.~\ref{eq:covfrac}). At the same time, our photo-ionization models predict that the ratio $N_{\rm{MgII,c}} / N_{\rm{HI,c}}$ decreases with increasing $N_{\rm{HI,c}}$, so that higher $N_{\rm{HI,c}}$ imply higher $\sigma_{\rm{los}}$ and/or higher $Z_{\rm{c}}$, explaining these two degeneracies. The velocity dispersion of the clouds seems rather well constrained and it is around $100$ km s$^{-1}$ (we discuss this further in Sect.~\ref{limitations}), while both the cloud column density and metallicity have very large uncertainties, implying that the choice of these two parameters does not strongly affect our outputs.

It is interesting to look at the recovered values of $\overline{N}_{\rm{HI,spaxel}}$: we find that the posterior distribution is in agreement with the previous estimates of the \HI{} column densities from B22 (see Table~\ref{tab:DLAmetricJ1527}), shown as a black vertical line in Fig.~\ref{fig:CornerPlotsJ1527}\footnote{We additionally run a Bayesian analysis fixing the value of the mean spaxel HI column density to the value found by B22 ($\log{(\overline{N}_{\rm{HI,spaxel}}/\rm{cm}^{-2})}=20.825\pm0.375$) and we obtain for the other free parameters very similar (and slightly narrower) posterior distributions to what presented here.}. However this quantity is very poorly constrained with rather large uncertainties, due mainly to strong degeneracies with both the column density and the metallicity of the clouds. We conclude that we recover the main finding of B22 that this absorber is a DLA, but also that our predictions of the total HI column density are very uncertain, as can be expected for a method that is based exclusively on a comparison with Mg\ion{ii} absorption lines. This also shows the limitations of the usual Mg\ion{ii} metrics (which are much more simplistic than our modeling) in selecting DLAs.

The final and most important result is given by the covering fraction $C_{\rm{frac}}$: our analysis prefers $\overline{C}_{\rm{frac}}<1$, with very tight constraints that are not degenerate with the other free parameters of the model. This result is consistent with our assumption that the DLA is composed by clouds and shows that these absorbers do not cover the entirety of the area traced by the gravitational arc. This is the first estimate of the structure of a DLA, which appears to be patchy on the scale of a MUSE spaxel (0.2\arcsec $\times$ 0.2\arcsec).

\subsubsection{J0033}
The right-hand side of Fig.~\ref{fig:CornerPlotsJ1527} shows the posterior distributions for the five free parameters in our analysis for the J0033 field. We can note that the intrinsic cloud kinematics, column density and metallicity ($\sigma_{\rm{los}}$, $N_{\rm{HI,c}}$ and $Z_{\rm{c}}$) have similar values and trends with respect to the J1527 field, although in this case they seem to have slightly lower uncertainties. 

The total neutral hydrogen column density, with a median value of the posterior distribution of about $10^{20.6}\ \rm{cm}^{-2}$, seems to point to the presence of a DLA even in this field. The posterior distribution of this parameter has however very large uncertainties (as for the J1527 case), hence this result will need further confirmation. Despite this, with our analysis we have a hint, more robust with respect to the usual Mg\ion{ii} metrics utilized in the literature, that this absorber is likely an extended DLA.

Very interestingly, the covering fraction of this extended absorber is again very well constrained and predicted to be lower than 1, with values that are even lower (but consistent within the uncertainties) with respect to the DLA in the J1527 field. This result seems hence to indicate that these two strong absorbers have similar properties and especially similar covering fractions, the most robust result of our Bayesian analysis. In both cases the gas distribution appears patchy, not covering completely the background extended source.

\section{Discussion}\label{discussionSection}

\subsection{Assumptions and limitations of model and data}\label{limitations}
The modeling framework presented in Sect.~\ref{modelSection} is idealized and relies on a number of assumptions, some of which we already discussed above. In the following we summarize and discuss the most important of these assumptions. 

First, we stress that the model is simplistic and it is not meant to represent a full 3D configuration of the absorbing gas. For this reason, we do not formulate predictions for the \HI\ and Mg\ion{ii} maps to directly compare with the real observational data (see e.g., Figs.~\ref{fig:DLAmapJ0033} and \ref{fig:DLAmapJ1527}). We instead assume that the clouds have the same kinematics, metallicity and column densities everywhere across the fields, not taking into account of possible (likely) inhomogeneities, which we only attribute to different spaxels detecting different numbers of clouds and having therefore different covering fractions and total column densities. While this assumption is clearly an oversimplification, the results of Sect.~\ref{ModelResults} show how, by simply comparing the total distributions of observational diagnostics like the \MgEW{} and the $D$-index, we can infer the general properties of the absorbing gas.

Another important approximation of the model is related to the cloud kinematics: we assume that the line-of-sight velocity distribution of the clouds is described by a Gaussian profile, which is most likely inaccurate. However, understanding the actual kinematics and dynamics of the absorbing gas is outside the scope of this work and the value of $\sigma_{\rm{los}}$ is simply used to determine the strength of the absorption: a non Gaussian kinematics would not strongly affect the outputs of our model (and therefore our main results and conclusions), as long as the average line-of-sight velocity dispersion is similar to what we found here. We also assume that the Gaussian distribution is centered at $0$ \kms{} in all the spaxels. This is justified by an ongoing kinematic analysis (Ledoux et al., in prep.) that shows that, at least for J0033, the bulk kinematics of the absorber is the same across the entire field. A variation of this center for different spaxels would anyway not impact our results, given that the strength of the absorption for each spaxel does not depend on the exact position of the velocity centroid but only on the velocity dispersion. However, we caution that the recovered value of $\sigma_{\rm{los}}$ depends slightly on the choice of the Doppler parameter $b$, dominated by the turbulent component $b_{\rm{turb}}$, which we fixed to $10$ km s$^{-1}$ in our fiducial model. We find that higher or lower values of $b_{\rm{turb}}$, while leaving the general conclusions of this study unchanged, would lead to respectively lower and higher values of $\sigma_{\rm{los}}$.

Finally, we inferred the column densities of Mg\ion{ii} using photo-ionized models \citep{ferland13}, while most of this gas is self-shielded and therefore in its neutral state. To investigate the impact of such choice on our results, we performed additional Bayesian analyses on the two fields assuming that the gas is entirely neutral and that the magnesium is all in the Mg\ion{ii} state. The Mg\ion{ii} cloud column density can therefore simply be obtained by assuming the magnesium solar abundance $\log(\rm{Mg}/\rm{H})_{\odot}=-4.47$ \citep{asplund09} and a value for the metallicity $Z_{\rm{c}}$. The results of such model are perfectly consistent, for both fields, with those shown in Fig.~\ref{fig:CornerPlotsJ1527}, so we conclude that the photo-ionization assumption does not have an impact on our main findings, as expected given the high column densities of the absorbers \citep[e.g.,][]{dey15}. We note that the photo-ionization assumption might have an influence on \FeMg{}, but we decided to exclude this diagnostic from our likelihood (Eq.~\ref{eq:likelihood}). Models where we also include a comparison between the observed and the predicted \FeMg{} distributions point towards the clouds having a low metallicity (median $\log Z_{\rm{c}}\approx -2.6$) and high HI column densities ($N_{\rm{HI,c}}\gtrsim 10^{19}\ \rm{cm}^{-2}$), but we choose to discard them, considering the uncertainties on the \FeMg{} diagnostics (see Sect.~\ref{sec:model}). Interestingly, even in this case the recovered covering fraction $\overline{C}_{\rm{frac}}$ remains consistent with the fiducial values reported in Sect.~\ref{ModelResults}.

The limitations of the data used may also impact the results of our modeling. As already mentioned in Sect.~\ref{metricsSec3}, our data might be contaminated by the galaxy light, resulting in an artificially weaker Mg\ion{ii} absorption in our spectra. In the modeling, higher \MgEW{} and $D$-indices could lead to slightly higher covering fractions with respect to what we obtained in Sect.~\ref{ModelResults}. Moreover, due to the seeing, the MUSE native spaxels (even after the smoothing described in Sect.~\ref{Musemaps}) are not spatially independent. The overall effect of the seeing is to smooth the properties of adjacent spaxels, so that we can expect that the overall impact on the distributions of \MgEW{} and $D$-indices is to make them narrower than what they would originally be. Given that our model predicts average quantities of the absorbing material across the two fields, using wider distributions would likely not change significantly our findings. We note that the distribution of measured EW and $D$-indices from spatially independent spaxels (i.e.,~spaxels separated by the point spread function of the observations, every 3.5 spaxels, which correspond to $\approx30$ and $\approx44$ spatially-independent measurements within the respective J0033 and J1527 fields) are consistent with the full spaxel distribution (with KS test p-values of $p\approx0.95$ for J1527 and $p\approx0.62$ for J0033). We therefore opt to use all the spaxels to improve the sampling of the EW and $D$-index distributions without significantly impacting our modeling analysis.

\begin{SCfigure*}[0.5][h!]
    \includegraphics[width=0.7\textwidth]{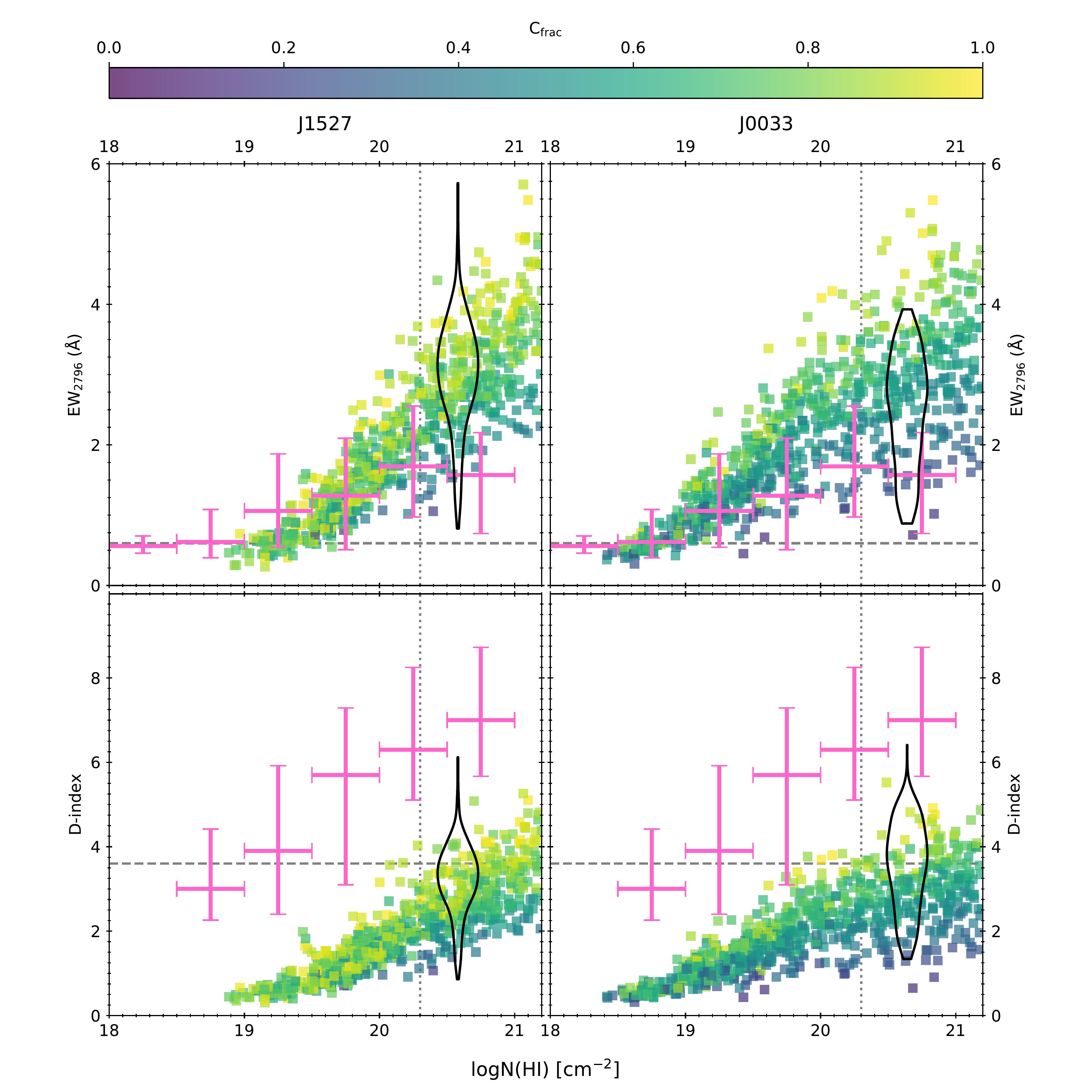}
    \caption{Model predictions of the DLA metrics  \MgEW{} (top panels) and $D$-index (bottom panels)  as a function of logN(\HI{}) in the two fields J1527 (left column) and J0033 (right column). The vertical dotted line denotes the DLA threshold logN(\HI)~$=20.3$~cm$^{-2}$ whilst the horizontal dashed line denote the minimum thresholds for a system to be considered a DLA for that metric; points in the top-right corner of each panel would be considered a DLA. The hollow violins show the distribution of \MgEW{} and $D$-index for spaxels (with S/N~$\geq2$ and \MgEW{} measured at $>2\sigma$ significance) within the field. The violins are arbitrarily centered at the median logN(\HI{}) predicted by the modeling, whilst their widths are arbitrarily set for display purposes and do not represent the error in logN(\HI{}). The pink errorbars show the median and 68$^{\rm th}$ percentile confidence interval of the literature measurements of single quasar sightlines \citep[taken from][]{Rao06,Ellison06,Berg17}. The literature values are binned in increments of 0.5 dex in logN(\HI{}). We note that the $D$-index measurements from the literature are obtained at a different spectral resolutions than the MUSE observations, and thus satisfy a different $D$-index threshold to be considered a DLA. These literature points are only shown to denote the dependence of $D$-index on logN(\HI{}) for individual quasar sightlines. The colored symbols represent the range of models within the 32nd and 68th percentiles of the posteriors of the four free parameters ($\sigma, \overline{C}_{\rm{frac}}, N_{\rm{HI,c}}, Z_{\rm{c}}$), and are color-coded by the covering fraction $C_{\rm{frac}}$. We stress that, while the literature data are obtained from point-source quasar sightlines, our model framework is designed to reproduce the gas absorption properties on extended sources.}
    \label{fig:metrics}
\end{SCfigure*}

\subsection{Evaluating the success of DLA metrics in individual spaxels}\label{DiscussionMetrics}

Figure~\ref{fig:metrics} encapsulates the dependence of \MgEW{} and $D$-index on logN(\HI{}) in the literature measurements taken from 1D spectra of quasar sightlines \citep[pink errorbars;][]{{Ellison06,Rao06,Berg17,Berg21}} and the best-fit models (colored symbols, obtained by sampling the posterior distributions of $\sigma_{\rm{los}}, \overline{C}_{\rm{frac}}, N_{\rm{HI,c}}, Z_{\rm{c}}$ within the 32nd and 68th percentiles) for both J1527 (left panels) and J0033 (right panels). The colored symbols represent a range of logN(\HI{}) that varies from the sub-DLA to the DLA regime. These distributions therefore are not intended to represent directly the distributions of the observed properties, but show instead the results of specific models as a function of the HI column density. For reference, the distribution of observed \MgEW{} and $D$-index for spaxels within the respective fields (i.e.,~spaxels in Figs.~\ref{fig:DLAmapJ0033} and \ref{fig:DLAmapJ1527} with S/N~$\geq2$ and \MgEW{} measured at $>2\sigma$ significance) are denoted by the hollow violins, and are centered on the median logN(\HI{}) predicted by the models. The general agreement of the violins with the model points in the top two rows of Fig.~\ref{fig:metrics} is a result of using these two distributions to constrain the models (i.e.,~Eq.~\ref{eq:likelihood}).  

The model is in general able to recover the bulk trends with logN(\HI{}) seen in the 1D quasar sightlines for the two metrics in Fig.~\ref{fig:metrics}, but there are some notable differences. First, the normalization of the $D$-indices (bottom row) is different between models and data, but this is expected because of the lower spectral resolution of our MUSE data with respect to the literature data (see also below). Second, the \MgEW{} (top row) tends to flatten in the 1D sightlines (in pink) for logN(\HI{})$\gtrsim20.3$ cm$^{-2}$, while it increases monotonically to larger values in the model, especially for J1527 (left column). This is due to the fact that these models are calibrated to our tomographic observations, which exhibit \MgEW{} significantly larger (see the position of the hollow violins) than the median values of the 1D data. To reproduce these high values, the \MgEW{} needs to increase at larger HI column densities, instead of reaching a plateau.

The third difference is instead given by the \MgEW{} scatter: whilst the scatter in the models appears similar (if not lower) to that of the 1D observations for sub-DLA HI column densities, the model scatter is instead much larger for logN(\HI{})$\gtrsim20.3$ cm$^{-2}$. A larger scatter is in principle not expected, given that in the modeling all clouds have the same metallicity (and density) within a single absorber (Sect.~\ref{sec:model}), while the mass-metallicity relation of absorbers \citep[e.g.,][]{Ledoux06,Neeleman13,Christensen14} is expected to influence the width of the observed Mg\ion{ii} absorption lines \citep[e.g.,][]{Ellison06,Rao06,Bouche08,Berg17} and thus to increase the scatter of the respective Mg\ion{ii} metrics. The large scatter in the model can be explained by the covering fraction, and a clear vertical gradient of covering fraction as a function of \MgEW{} (top row) or $D$-index (bottom row) for a given logN(\HI{}) is indeed visible in Fig.~\ref{fig:metrics}. As a result, it is clear that using the same Mg\ion{ii} metric thresholds from 1D quasar sightlines on 2D extended spaxels may not be so straightforward to interpret. Given the small pencil-beam nature of quasar sightlines, the covering fraction across the quasar would only vary if the angular sizes of the gas clouds comprising the absorber are much smaller than the angular size of the continuum emission from the background quasar, which seems unlikely. However covering fraction should impact extended sources, where the coherence length of Mg\ion{ii} absorption is expected to be $\sim5$~kpc \citep{Afruni23} with cloud sizes of $\sim0.5$~kpc \citep[for both low, e.g., Mg\ion{ii}, and intermediate, e.g., C\ion{iv}, ionization gas;][]{Faerman23,lopez24}. For reference, as already mentioned, the range in spaxel sizes in both fields is 0.07 and 1.4 kpc along a side (after delensing).

\begin{table}
    \centering
    \caption{Fraction of modeled spaxels with logN(\HI{})~$\geq20.3$ (f$_{\rm DLA}$) for both J0033 and J1527 fields for a given threshold for \MgEW{} and $D$-index.}
    \label{tab:metricSuccess}
    \begin{tabular}{lcc}
    \hline
    \hline
f$_{\rm DLA}$ & \MgEW{} threshold & $D$-index threshold\\
\% & \AA & \\
\hline
\multicolumn{3}{c}{J0033}\\
30 & 0.31 & 0.32 \\
40 & 0.70 & 0.68 \\
50 & 1.42 & 1.35 \\
60 & 2.01 & 1.88\\
70 & 2.46 & 2.33\\
80 & 2.86 & 2.74\\
90 & 3.48 & 3.37\\
100 & 4.19 & 3.84\\
\hline
\multicolumn{3}{c}{J1527}\\
30 & 0.27 & 0.31 \\
40 & 0.27 & 0.31 \\
50 & 0.63 & 0.63 \\
60 & 1.15 & 1.06\\
70 & 1.52 & 1.46\\
80 & 1.86 & 1.80\\
90 & 2.28 & 2.11\\
100 & 3.69 & 4.03\\
\hline
\end{tabular}

\end{table}

Another way to assess the success of the DLA Mg\ion{ii} metrics is to look at the fraction of the models that produce a DLA (f$_{\rm DLA}$) for a given metric threshold. Table~\ref{tab:metricSuccess} provides the required threshold for \MgEW{} and $D$-index to produce a given f$_{\rm DLA}$ for the models of both fields. We note that f$_{\rm DLA}$ is $\approx 20$\% and $\approx40$\% using the typically adopted \MgEW{} thresholds of 0.3\AA{} and 0.6\AA{} in the literature, whilst f$_{\rm DLA}\approx90$\% \citep[i.e.,~similar to the purity found in][]{Ellison06} using the $D$-index cut expected for MUSE-like resolution ($\approx3.6$). The \MgEW{} threshold obtained by the simulations is within a factor of two of the equivalent thresholds required to reproduce the same f$_{\rm DLA}$ seen in the high redshift ($2<z<4.5$) XQ-100 survey \citep{Berg17, Berg21}, while the lower-redshift ($z<1.65$) sample from \cite{Rao06} requires a consistently high \MgEW{} threshold of $\gtrsim2.7$ in order to reproduce the same f$_{\rm DLA}$. Whilst the scatter in the relations of Fig.~\ref{fig:metrics} is larger in the modeled 2D spaxels in comparison to the 1D literature quasar sightlines, it appears that the purity of the sample (i.e., $f_{\rm{DLA}}$) for a given \MgEW{} threshold is roughly the same between the two types of background sources. The fact that our modeling is able to reproduce the observed trends and statistics for pencil-beam quasar sightlines validates that we are able to estimate the \HI\ column density (albeit with large uncertainties) and the gas covering fraction based on Mg\ion{ii} absorption towards extended background sources.

\subsection{The nature of the absorbers}
The results of Sect.~\ref{modelSection} can be used to understand what is the nature of the two strong absorbers detected in the J1527 and J0033 fields and what is their potential role in the evolution of the galaxies with which they are associated.

Looking at the projected distances in the absorber planes reported in Table~\ref{tab:regions}, we note that the two strong absorbers (both potentially DLAs) extend up to distances of a few tens of kpc. Compared to the typical sizes of galactic discs at redshift $z>1$ \citep[e.g.,][]{shibuya15}, these absorbers seem to be significantly more extended. This indicates that this gas is likely more representative of the CGM, rather than of high-redshift rotating interstellar gas discs \citep[see][]{neeleman20,kaur24}.

By summing the areas in the absorber plane (using the same lens models mentioned in Sect.~\ref{observations}) of all the MUSE spaxels, we can calculate the total area subtended by the extended source in the J1527 and J0033 fields, respectively $353$ kpc$^2$ and 350 kpc$^2$. Assuming a mean HI column density across the field equal to the median value of the two posterior distributions of $\overline{N}_{\rm{HI,spaxel}}$ (see Fig.~\ref{fig:CornerPlotsJ1527})\footnote{Being the `observed' HI column density, $\overline{N}_{\rm{HI,spaxel}}$ already accounts for the covering fraction (Eq.~\ref{eq:covfrac}).} we then obtain $M_{\rm{HI}}\gtrsim 2\times10^9\ M_{\odot}$ for J1527, consistent with the previous results of B22, and $M_{\rm{HI}}\gtrsim 1.2\times10^9\ M_{\odot}$ for J0033. We stress that these two values represent strictly lower limits, given that the extension of the absorber could be larger than the extension of the background arc.

Our picture is consistent with the recent results of \cite{stern21}, who using the FIRE simulations find that a predominantly neutral inner CGM, potentially giving rise to DLAs, is present in halos at $z>1$ with virial masses of about $10^{11}\ M_{\odot}$, where the cooling time of the hot gas is shorter than the free-fall time of the system. These authors find that in these halos the neutral gas can extend with large covering fractions ($0.5<C_{\rm{frac}}<1.0$) up to distances of tens of kpc from the host galaxy, in agreement with our findings. Moreover, in their simulations the cold gas is on average inflowing towards the galaxy (even though with relatively low infall velocities). If this scenario is correct, and we are detecting such material with our observations, this large amount of cold, neutral gas ($M_{\rm{HI}}\gtrsim 10^9\ M_{\odot}$) would account for a substantial fraction of the baryons within the halo and would therefore represent the main source of fuel for the future star formation within the galaxy.

We conclude by emphasizing that, except for the above speculations, here we do not attempt to draw a full dynamical picture for this gas, especially since we can not determine which galaxies are responsible for these absorbers: for J1527, the MUSE spectral coverage does not allow us to detect galaxies in emission at the redshift of the absorber ($z_{\rm{abs}}\simeq2.06$), while in the J0033 field we detect multiple galaxies (at $z_{\rm{abs}}\simeq1.17$) whose CGM could contribute to the absorption (Ledoux et al., in prep.). Disentangling whether this gas is a signature of galactic winds \citep[e.g.,][]{schroetter19,schneider20,FernandezFigueroa22}, recycling material, or cold cosmological accretion \citep[e.g.,][]{vandevoort11,bouche13,theuns21} is therefore outside the scope of the present study and is left for future work.

\section{Summary and conclusions}\label{conclusions}

In this paper, we analyzed two gravitational arc fields with known strong Mg\ion{ii} absorption in order to predict the presence of a \HI{}-rich absorber (such as a DLA) and determine the spatial extent of the absorber. For the first time, we created 2D maps of strong Mg\ion{ii} absorbers using three Mg\ion{ii} metrics (\MgEW, $D$-index, and \FeMg) typically used in the literature for identifying DLAs \citep{Rao00,Ellison06}. These maps suggest that both gravitational arcs probe subDLAs or DLAs, and are extended over areas of $\approx350$ kpc$^{2}$. In particular, one system (J0033) shows $\leq50$~\kms{} variations in the redshift of the observed Mg\ion{ii} absorption across $\approx50$~kpc of separation between sources, suggesting the same bulk gas is extended over such very large areas. The Mg\ion{ii} metrics for the other field (J1527) suggest the presence of a DLA, which has been confirmed by KCWI data of the same system \citep{Bordoloi22}. These Mg\ion{ii} metrics can be successful for implying the strength of \HI{} absorption along extended emission sources.

In order to quantify the success of these metrics, as well as provide an estimate of the \HI{} column density towards the J0033 field, we developed a simple toy model in order to evaluate the robustness of the three Mg\ion{ii} metrics used to identify DLAs in 1D quasar sightlines. The results from our toy model, which assumes a series of clouds in front of an extended background source, suggest that the typical covering fraction of gas clouds in front of the two gravitational arcs is $\approx60 - 80$~\%. The resulting purity of these metrics in selecting DLA column densities in front of extended sources is similar to what has been observed in the literature for 1D quasar sightlines \citep{Rao06,Ellison06,Berg17}.  The model also demonstrates that the covering fraction of gas can also influence both the \MgEW{} and $D$-index metrics measured. As a result, choosing a threshold for these two metrics to identify DLAs towards extended sources is not straightforward. For \MgEW{}, the threshold can depend on metallicity \citep{Rao06,Berg17} and covering fraction. Whilst the $D$-index threshold depends on the spectral resolution of the observations \citep{Ellison06} and covering fraction, we suspect that the covering fraction should not influence point-source, quasar observations as the typical cloud size is likely much larger than the background source. The typical spaxel size can vary between 0.07 and 1.4 kpc in our two fields after delensing. These results provide a cautionary tale of using Mg\ion{ii} metrics to identify DLAs in front of extended background sources, despite the success at identifying the DLAs in both of these systems.

Despite the difficulty of using the Mg\ion{ii} metrics, our model is able to predict for the first time the covering fraction of gas that gives rise to DLAs, providing unprecedented insight on the structure of this medium. Moreover, for the J0033 field we were able to estimate the previously unknown logN(\HI{}) of the absorber, logN(\HI{})~$=20.6 \pm 0.3$, consistent with a DLA. 
Using the area of the delensed arc, we predict a lower limit of the total \HI{} gas mass to be $\gtrsim10^{9}$ M$_{\odot}$ in both absorbers.  We speculate that both absorbers are part of a neutral inner CGM and that, given the large amount of mass, they could be an essential source of fuel for future star formation in the host galaxies.

\section*{Data Availability}
The median metal line profiles obtained for the different regions of the J1527 and J0033 fields are available online respectively in Appendix B (\url{https://zenodo.org/records/14225814}) and Appendix C (\url{https://zenodo.org/records/14225839}).

\vspace{.2cm}

\begin{acknowledgements}
We thank both Rongmon Bordoloi and Sara Ellison for respectively providing us with the aperture information used in B22 and the $D$-index data from \cite{Ellison06}. We also are grateful for Keren Sharon for creating the lens models for both of these systems. Finally, we thank the anonymous referee for a constructive and thorough report. S.L. acknowledges support by FONDECYT grant 1231187. This work is based on observations collected at the European Organisation for Astronomical Research in the Southern Hemisphere under ESO programme(s) 098.A-0459(A) and 0103.A-0485(B).
    
\end{acknowledgements}

\bibliographystyle{aa}
\bibliography{references}

\begin{appendix}

\section{Continuum fitting}
\label{sec:contfit}

In order to continuum fit the spectra from every spaxel of the MUSE cubes and to propagate continuum fitting uncertainties within a Monte Carlo framework, we used a simple sigma-clipping method to automatically determine the continuum of a given spectrum.  In summary, the continuum fitting method uses the \textsc{astropy} \citep{Astropy} sigma clipping algorithm within a window of the spectrum of width $1000\AA{}$ (800 pixels). The algorithm determines the mean flux ($\Bar{F}$) within the window while rejecting all pixels with flux outside of the confidence interval $\Bar{F}-n_{\rm{low}}\sigma$ and $\Bar{F}+n_{\rm{high}}\sigma$ (where $\sigma$ is the standard deviation of all flux measurements within the window). The continuum is built-up by repeating this across the entire spectrum by sliding the window in increments  of $100\AA{}$ (80 pixels). Once completed, a spline is fit to the points generated from the sliding window (centered in the window for each step), and the spline is then smoothed with a top-hat function of width of 300\AA{} to remove strong variations in the flux from regions of the spectra with a low S/N. We note that the continuum of the background sources are free of emission lines and smooth, allowing us to use a large window and improve the statistics used in the sigma-clipping method. While we fix $n_{\rm{high}}=1.0$ in the sigma clipping algorithm, we found the optimal choice of $n_{\rm{low}}$ to distinguish between noise and absorption depends on the average S/N of the spectrum. We therefore adjust $n_{\rm{low}}$ across each spectrum. Table~\ref{tab:snrn} shows the optimized $n_{\rm{low}}$ adopted for a given S/N (per pixel) to exclude absorption in the continuum fitting process.

\begin{table}
    \centering
    \caption{Lower sigma clipping thresholds for average S/N of spectrum}
    \label{tab:snrn}
    \begin{tabular}{lc}
        \hline
        \hline
        S/N & $n_{\rm{low}}$ \\
        (pixel$^{-1}$) & ($\sigma$) \\
        \hline
        0.1 & 0.98 \\
        0.2 & 0.94 \\
        0.3 & 0.93 \\
        0.4 & 0.90 \\
        0.5 & 0.88 \\
        0.8 & 0.83 \\
        1.0 & 0.78 \\
        1.2 & 0.75 \\
        1.5 & 0.71 \\
        2.0 & 0.66 \\
        2.5 & 0.64 \\
        3.0 & 0.63 \\
        3.5 & 0.64 \\
        4.0 & 0.64 \\
        4.5 & 0.64 \\
        5.0 & 0.65 \\
        7.0 & 0.65 \\
        10.0 & 0.62 \\
        \hline
    \end{tabular}

\end{table}

\end{appendix}

\end{document}